\pdfoutput=1
\documentclass[11pt,a4paper]{article}
\usepackage{jinstpub}
\usepackage{graphics}
\usepackage{float}
\usepackage{url}
\usepackage{subcaption}

\usepackage[font=small,labelfont=bf]{caption}
\usepackage{subcaption}

\title{New approaches to particle induced prompt gamma imaging }

\author[a]{Mythra Varun Nemallapudi\note{Corresponding
author,}} 
\author[a,b]{Atiq Ur Rahman,} 
\author[a]{Vanny Marantha Sihotang,}
\author[b]{ Thien Cong Tran,}
\author[c]{ Hsin-Hon Lin,}
\author[a]{Chih-Hsun Lin,}
\author[a]{ Ming-Lee Chu,}
\author[b]{Augustine Ei-Fong Chen,}
\author[a]{ Shih-Chang Lee}

 \affiliation[a]{Institute of Physics Academia Sinica,\\
No 128, Academia Rd, Nangang district, Taipei, Taiwan}
\affiliation[b]{Department of Physics, National Central University,\\
No 300, Zhongda Rd, Zhongli district, Taoyuan, 320}
\affiliation[c]{Institute of Radiological Research, Chang Gung University,\\
No 259, Wenhua 1$^{st}$ Rd, Guishan district, Taoyuan, 333}

\emailAdd{mythravarun01@gmail.com}

\abstract{ The distribution of the prompt gamma emissions  induced by proton interactions in a target material carries information concerning the proton range and target composition. We propose multiple approaches for prompt gamma imaging o verify proton range or to identify the material composition based on different detection scenarios. The first approach is based on knife-edge collimator utilized in tandem with a compact PET type detector  in order to verify  proton range by measuring the depth distribution of prompt gamma relying on Compton scattered events. A second approach aims at unambiguously identifying the presence of protons in a region of interest by introducing a bio-compatible fiducial marker in the patient and measuring the characteristic prompt gamma of the marker against background material.  A third approach illustrates a novel design for position sensitive gamma imaging with a millimeter level adjustable position resolution with a high detection efficiency which is a step towards 3D prompt gamma imaging, useful to better understand the target composition in addition to range verification. Simulations  of the designs were performed using GATE/Geant4 Monte Carlo framework. Experimental tests for some cases were conducted at the proton facilities in INER and Chang Gung Memorial hospital. Using the knife-edge PET approach a range verification of $\pm$0.7 mm can be achieved for shifts within 1 cm near the tumor region and up to  $\pm$ 4 mm for shifts within a 4 cm window.  For the fiducial marker approach we identify 984 keV as the dominant prompt gamma and show that for the 50\% relative PG intensity, the R80 position falls within a 3 mm thick $\mathrm{^{48}}$Ti  marker. Using the position sensitive gamma imaging approach we present a feasible design to achieve spatial resolution values of 2.6 mm with a detection efficiency of $\mathrm{5.4\times 10^{-6}}$  at 6.1 MeV and upto eight depth positions in a single run. The relative advantage of these methods and the challenges in the implementation are discussed. 
}

\begin{document}

\maketitle
\section{Introduction}\label{section_Introduction}
High energy protons interact with a target material through various processes resulting in the loss of proton energy as they traverse the medium and along with it  generate secondary particles  and isotopes (which produce prompt and delayed gamma) through  nuclear interactions with the target material. The physics of proton interactions are described in \cite{ProtonPhysics}. 

\begin{figure}[h] 
\begin{minipage}[h]{0.36\textwidth}
{\centering\includegraphics[width=\textwidth]{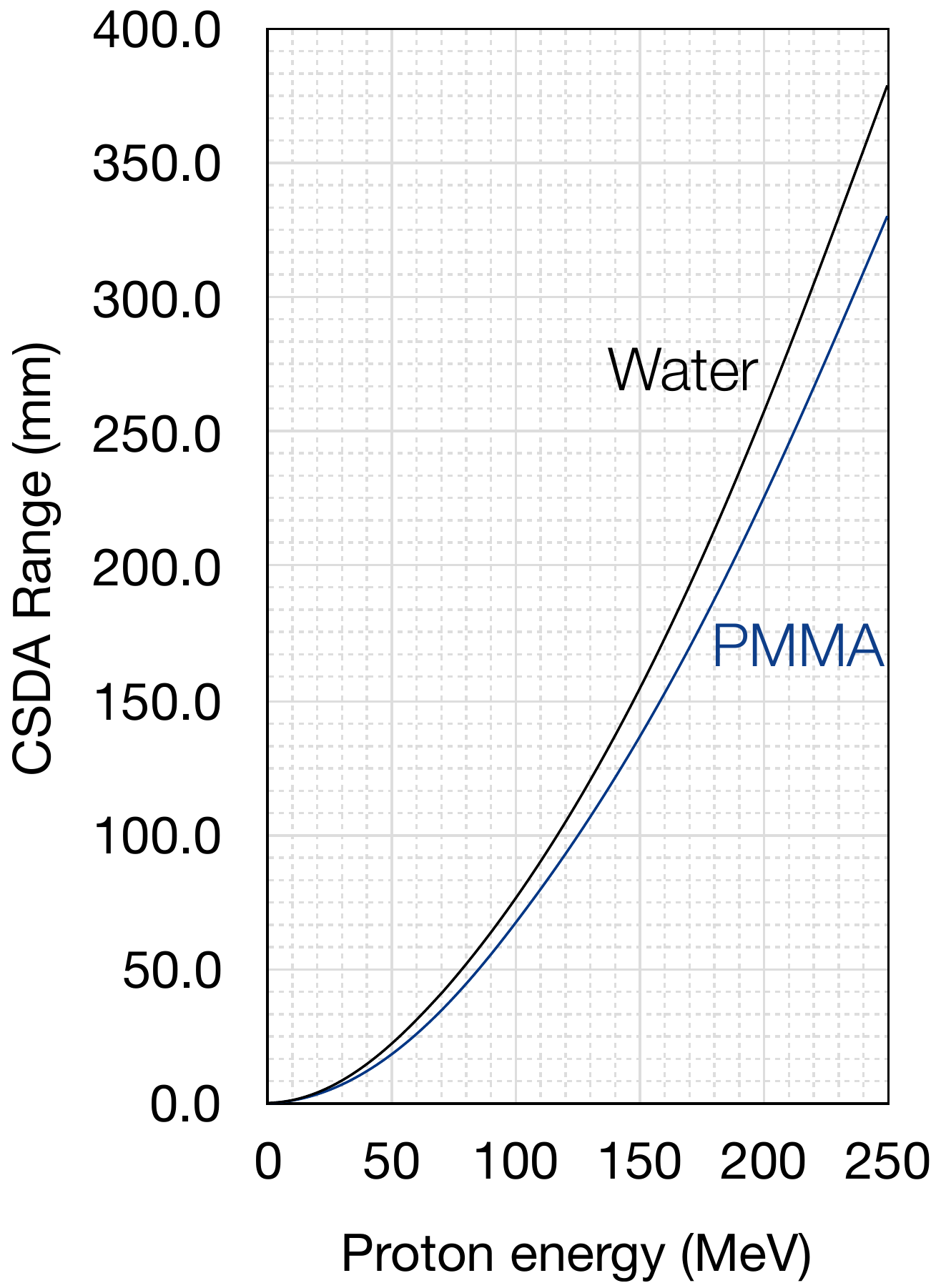}}
\end{minipage}
\hspace{0.00\linewidth}
\begin{minipage}[h]{0.64\textwidth}
{\centering\includegraphics[width=\textwidth]{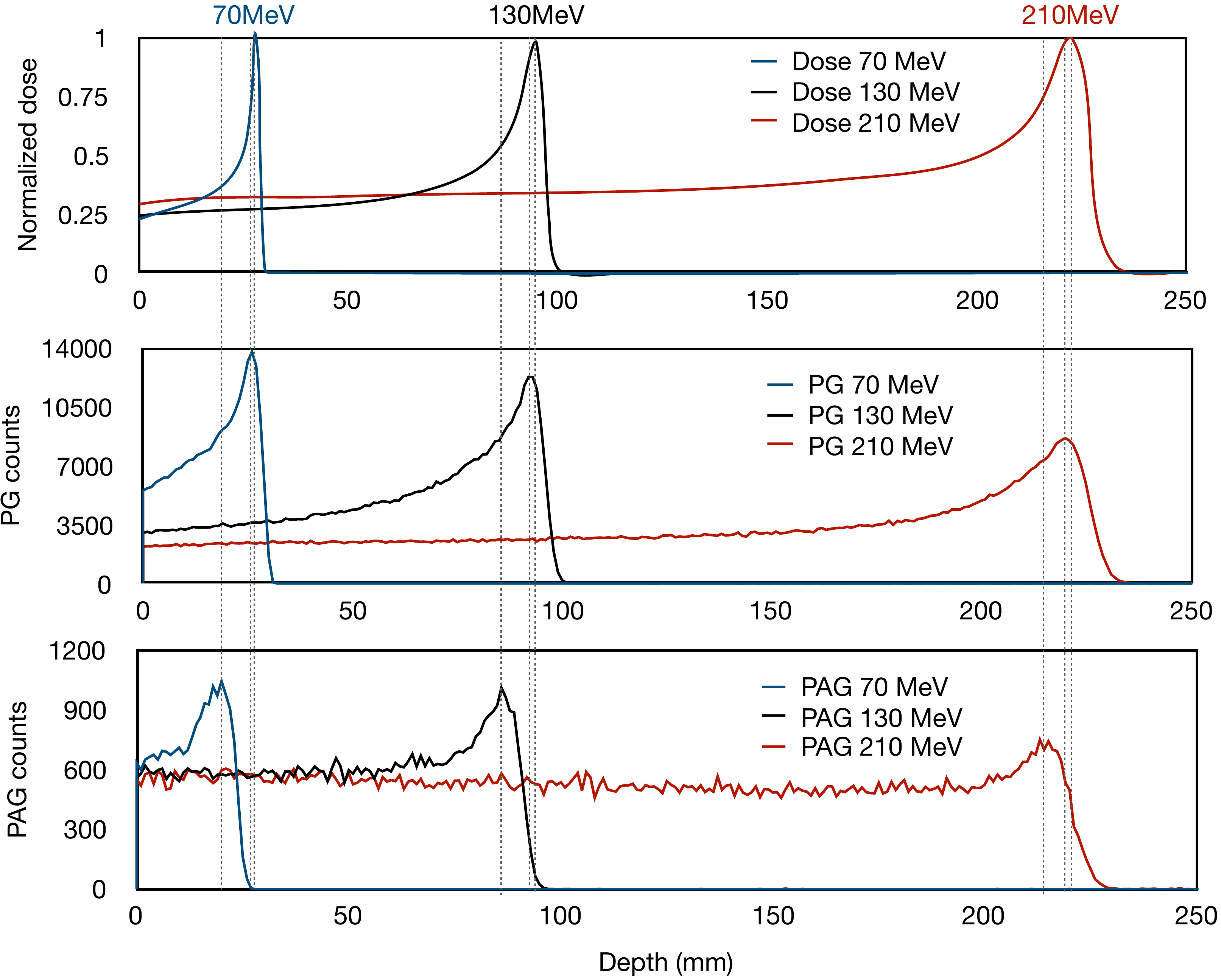}}
\end{minipage}
\caption{Left: Proton range with a continuous slowing down approximation vs Energy in water and PMMA, the data was obtained from PTSTAR tables\cite{PTSTAR}. Right: Geant4 simulated depth distributions of normalized dose (top), prompt gamma (middle) and positron annihilated gamma isotopes (bottom) for protons of 70, 130 and 210MeV in a PMMA target. $\mathrm{10^{7}}$ protons were irradiated for the simulation.}
\label{fig_Intro_Depth}
\end{figure}

Secondary isotopes are created in accordance to their energy dependent cross sections. The excited nuclei subsequently undergoing fast de-excitation yield prompt gamma (PG) such  as $\mathrm{^{12}C^*}$ and $\mathrm{^{16}O^*}$ in materials rich in carbon and oxygen. Some of the secondary isotopes ($\mathrm{^{12}C(p,pn) ^{11}C}$, $\mathrm{^{16}O(p,pn)^{15}O}$) decay via positron emission to produce positron annihilation gamma (PAG) which comprises a coincident pair of 511 keV gamma.
The energy dependent production cross sections increase as the proton energy decreases and vanish completely at the threshold proton energy required for the production. As seen in figure \ref{fig_Intro_Depth}, in water, the peak of the secondary gamma emitting isotope distribution lags behind the peak of the depth dose distribution (Bragg Peak) from 3.1 mm - 3.3 mm for the dominant PG isotopes and 7.2 mm - 8.4 mm for the dominant PAG isotopes depending on the incident proton energy (70 MeV  to 210 MeV)  and the production threshold energy. The small range of deviation among the differences in PG peak position with respect to the dose position (similarly for PAG) over several proton energies enables us, in principle,  to utilize this information to verify the range at a sub-mm level for a normal treatment scenario.

Since the gamma production cross sections depend on proton energy and material composition, imaging proton induced gamma can offer information useful a) to verify the range of the incident protons and b) to map out the material composition \cite{RKL1979}, \cite{Dyer1981}. The risk posed to sensitive organs during proton therapy is mitigated in routine clinical usage by introducing safety margins in the treatment planning (2.5-3\% proton range + 2-3 mm) and by choosing safer directions of delivery \cite{Paganetti_Range}. However verifying the proton range through direct or  indirect methods can ensure that the dose delivered conforms to the treatment plan, and also allows more aggressive treatment planning in the proximity of sensitive organs \cite{Krimmer2018}. 


Several groups worked on the range-verification problem by addressing  the above mentioned challenges by using different approaches. Most of them use some form of PG/PAG detector with or without collimation to record the count distribution on the detector with or without energy resolving, this information is used to reconstruct the original distribution in the target and compare it with the expected value.

PAG provides an unambiguous way to measure the depth distribution of the positron activity along the target material. The detected coincidence-pairs of 511 keV enable the reconstruction of positron activity which can be compared with the expected distribution to predict the range shifts.  Starting from Maccabee et al in 1969 \cite{Maccabee1969},  several groups have investigated this approach  \cite{Bennett1975}, \cite{Enghardt1992}, \cite{Parodi2001}, \cite{Parodi2007}, \cite{Nishio2010}, \cite{Shao2014}, \cite{Piliero2016}, \cite{Binet2018}, \cite{Crespo2005}, \cite{Buitenhuis2017} ,  \cite{Tashima2020}. Some of the salient aspects of the PAG imaging approach, along with our PET design (proposed to be used in  tandem with a knife-edge as discussed in section \ref{section_KE}) and initial beam tests on a smaller scale can be found in our previous work \cite{Nemallapudi2021}. Although the existence of 511 keV pairs makes the localization relatively  straightforward, imaging PAG has limitations posed by a smaller production number, delayed imaging that can be blurred due to washout effects, positioning errors if the patient is to be moved, and strain on the beam time if the patient is imaged on the same couch. As of the year 2021 are 112 proton therapy facilities around the world and over half of them are based on cyclotron \cite{PTCOG2021}, and although in-beam PET imaging has been previously demonstrated, there is a need for other practical solutions for range verification. 

Detecting PG distributions to correlate with particle range addresses some of the above described limitations of PAG imaging and this was first proposed by \cite{Stichelbaut2003}. The detector design required for PG imaging is different from PET - larger crystals, smaller number of channels, thick collimators, shielding to protect the detectors from radiation damage etc. 

%

Prompt gamma detectors have been developed by various groups based on different approaches. 
Slit-collimator based designs restrict the gamma detection to a narrow target region using thick collimators to form a parallel slit. When the detector uses a high energy resolution crystal, the spectroscopic information can be obtained. The systems reported by \cite{Verburg2018} based on the ideas first presented in \cite{Verburg2014} utilize eight modules of cylindrical LaBr$_3$ crystals (length 76.2 mm, diameter 50.8 mm) arranged in two columns behind a slit collimator (127 mm thick, 102 mm wide, 12.7 mm slit width). Absolute proton range was measured to 1.1 mm at a 95\% confidence interval.    

The knife-edge collimators and multi-slit collimator based detectors designs can be seen in  \cite{Perali2014}, \cite{Smeets2016}, \cite{Lin2017} and \cite{Park2019}. A knife-edge collimator is an alternative to a  slit collimator and results in a higher detection efficiency and a greater positional range working effectively like a pin-hole camera that inverts the detected image from the source distribution along the depth axis.  The design presented by \cite{Perali2014} utilizes 40 LYSO crystals of $\mathrm{4\ mm \times \ 30\ mm\times \ 100 \ mm }$ size located 200 mm from the center of a knife-edge collimator made of tungsten with a slit width of 6 mm and a slit angle of  63$^o$. Based on the provided values, this system weights 16 kg, owing mainly to the collimator dimensions and needs to be mounted on a  dedicated station.

Compton camera approaches were presented by \cite{Krimmer2015}, \cite{Solevi2016}, \cite{Draeger2018} and \cite{Yao2019}. A range shift prediction to within 2 mm was shown by a system presented by \cite{Draeger2018} based on a partial implementation of a multiple detection stage Compton imager J-Polaris weighing 4 kg.  The arrangement of the camera extends over 100 cm and seems to be geometrically challenging for a clinical implementation. 

Spectroscopic detectors presented by \cite{Panaino2019} aim to identify the energy information of the prompt gamma from special modes of emission such as the triple coincidence from the $\mathrm{^{16}O^*}$ nucleus. Based on a 4-PI coverage and close proximity (8 cm from beam axis)  of the front face of the detector to the target they illustrate that range could be determined with an uncertainty below 7 mm within a 68\% confidence interval for $\mathrm{10^{8}}$ protons. The design is mechanically unrealistic for integration in a therapy setting.
The detector developed by \cite{Ready2016} et al. utilizes a collection  of several knife edges in a specialized arrangement resulting in a unique image on the detector similar to a coded aperture. Using 50 MeV proton beam the range shifts were measured to within 1 mm (2$\sigma$) for $\mathrm{1.7\pm0.8\times 10^{8}}$ protons. Smaller unit crystal size of this design limits the ability to image the distribution of multiple PG energies and estimate the material composition.


The detectors described above were designed with range verification as a goal. In addition to range verification, in order to identify the material composition we need a good localization of the gamma, an acceptable energy resolution and a high detection efficiency.  This can have interesting applications in the non-destructive analysis of arbitrary targets such as mineral samples or living biological tissue.
Conventional imaging techniques are not well suited for this task. Techniques such  as mass spectroscopy and chemical analysis obtain the elemental fractions in homogeneous samples and do so at a surface level \cite{MassSpectroscopy}. X-ray diffraction requires a regular arrangement of the elements within the sample to give rise to diffraction peaks and does not provide sharp diffraction peaks when using amorphous samples.  Computed tomography although capable of achieving high resolution images, is based on X-ray attenuation and does not provide an unambiguous identification of the elemental composition.  

The ability to image the elements using proton induced  gamma imaging (PIGI) can provide information on the materials in a way that is different from the existing methods. Higher proton energies are needed to probe bulk samples non destructively using a  method that can image prompt gamma as a function of beam position while also offering positional information.


%


\paragraph{Goal of this paper}
In this work we present three new approaches with a singular aim of obtaining useful information from proton-induced gamma in a practical and efficient manner. Each approach takes advantage of a specific aspect of the detection scheme in view of the desired information. We discuss the important considerations for designing a prompt gamma detector in section \ref{section_DetectorDesign}. In section \ref{section_KE} the first approach, a knife-edge collimator in tandem with PET, is presented as a feasible design to utilize partial energy depositions in smaller crystals. A simulation study for range verification is presented involving compact detectors which can be better integrated in a therapy facility. In section \ref{section_FM} we present our investigations on the idea of using a marker with a unique gamma signature to verify the proton range as our second approach. In section \ref{section_3DPG} we propose our third approach, a novel way of building a multi channel detector based on a fundamental block using a basic slit collimator to obtain a depth distribution of individual PG while maintaining a high detection efficiency and spatial resolution. This approach can lead to 3D Prompt Gamma imaging (3DPG).
These approaches are aimed to overcome some of the limitations in the existing detectors which limit their application to a smaller class of problems. 

\section{Materials and Methods \label{Setup}} \label{section_Materials}

\subsection{Simulation}
The Monte Carlo (MC) package GATE (version9.0) \cite{GATE} which is based on Geant4 (version 10.5.1)  simulation toolkit \cite{Geant4} has been used to perform simulations relevant for our prompt gamma detector  development. The physics list QGSP\_ BIC\_ HP\_ EMY  \cite{Geant4-physicslist} accounts for the processes involved in the proton beam interactions with the target material, the process of dose deposition within the target, the generation of secondary particles due to nuclear interactions, the interaction of the secondary particles within the phantom material and subsequently with the detector system involving mechanical collimators and scintillating crystals. Binary cascade and pre-compound model are used to implement the nuclear interactions depending on the proton energy, and photon evaporation model describes the decay of the excited nuclei to produce gamma. The optical sensor is not modeled in our simulations, and the reported counts and energy are based on the gamma energy deposited in the crystals. The energy dependent resolution is modeled for the KE-PET in section \ref{section_KE} but not for the 3DPG detector in section \ref{section_3DPG}. 

The simulated distributions of the 984 keV gamma in section \ref{section_FM} and the carbon and oxygen PG in section \ref{section_3DPG} are model dependent and not always  in agreement with the experimental values, however this does not significantly impact the approaches described in this work and the designs can be further optimized.

\subsection{Experimental setup for proton-target interactions}
\label{section_ExperimentalSetup}
The experiments described in this  work have been performed using a coaxial HPGe detector (Canberra GC3022), although the crystal dimensions are not specified, \cite{Jeskovsky2019} describe it to be a cylinder of 5.73 cm diameter and 5.75 cm height. The detector operated at 2500 V and the energy spectrum was collected using  CAEN DT5770 MCA. The energy resolution of the HPGe GC3022 at 661 keV was measured to be 0.81\%. The deteriorated energy resolution can be attributed to radiation damage, increased pressure in the cryostat, and other aging effects over 20+ years of usage; nevertheless all important gamma lines for our  tests were clearly identified.  Lead and tungsten collimators were used to shield the detector where necessary. The cyclotron at INER was used as a proton source for 30 MeV protons in the experiment with Ti target mounted on a water phantom described in section \ref{section_FM}. A 230 MeV cyclotron installed at the CGMH proton facility delivered a pencil  beam between the energies of 70 MeV (beam width of 4.7 mm FWHM) and 90 MeV (beam width of 7.4 mm FWHM) for a beam energy scan on a Ti marker with water phantom described in section \ref{section_FM}.


\section{Considerations for detector design}
\label{section_DetectorDesign}
\begin{figure}[h] 
\centering
	\begin{subfigure}[b]{0.5\textwidth} 
	\centering
	\includegraphics[width=0.99\textwidth]{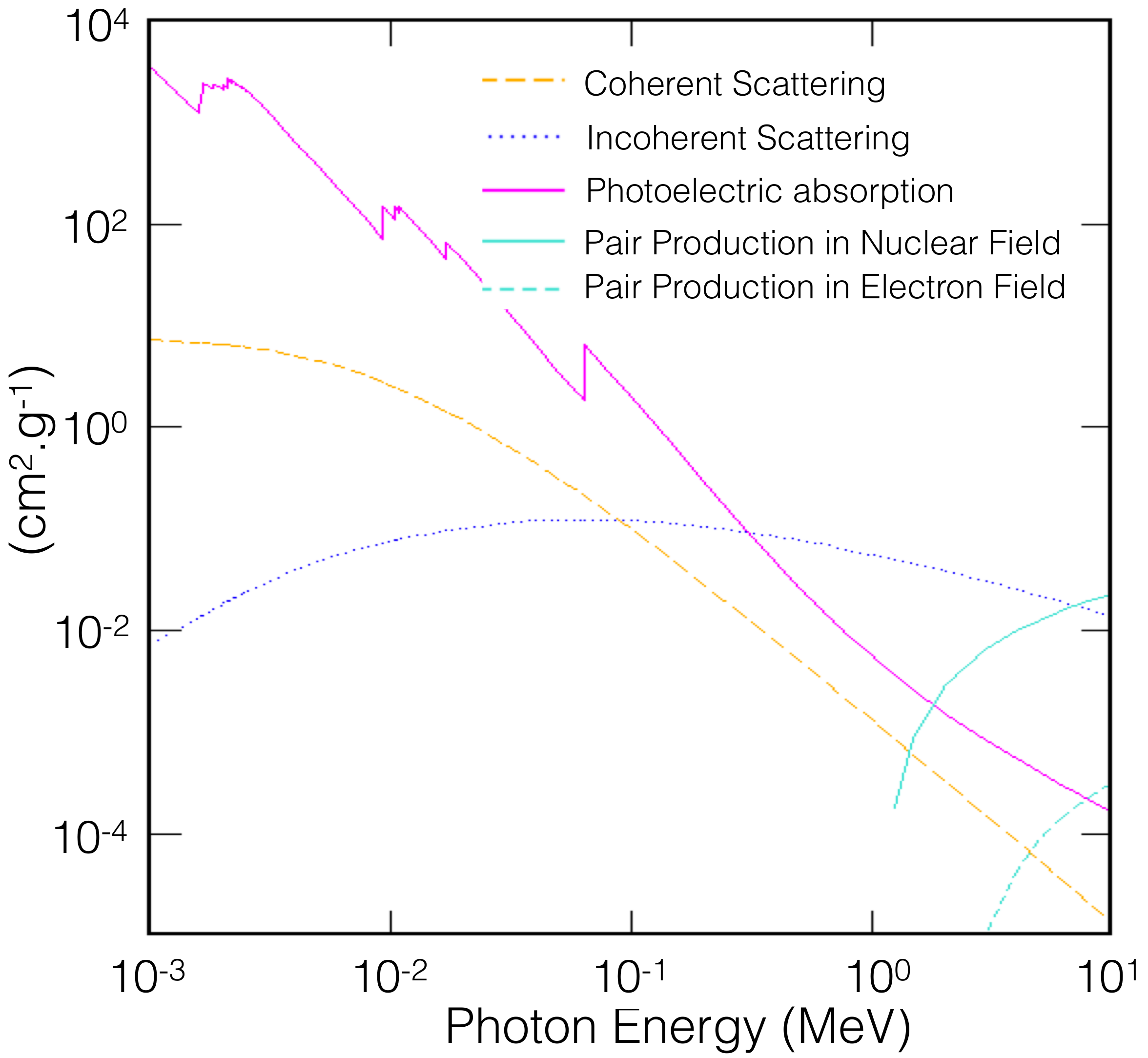}
	\caption{}
	\end{subfigure}
	\hfill
    \hspace{0.00\textwidth} 
	\begin{subfigure}[b]{0.99\textwidth} 
	\centering
	\includegraphics[width=\textwidth]{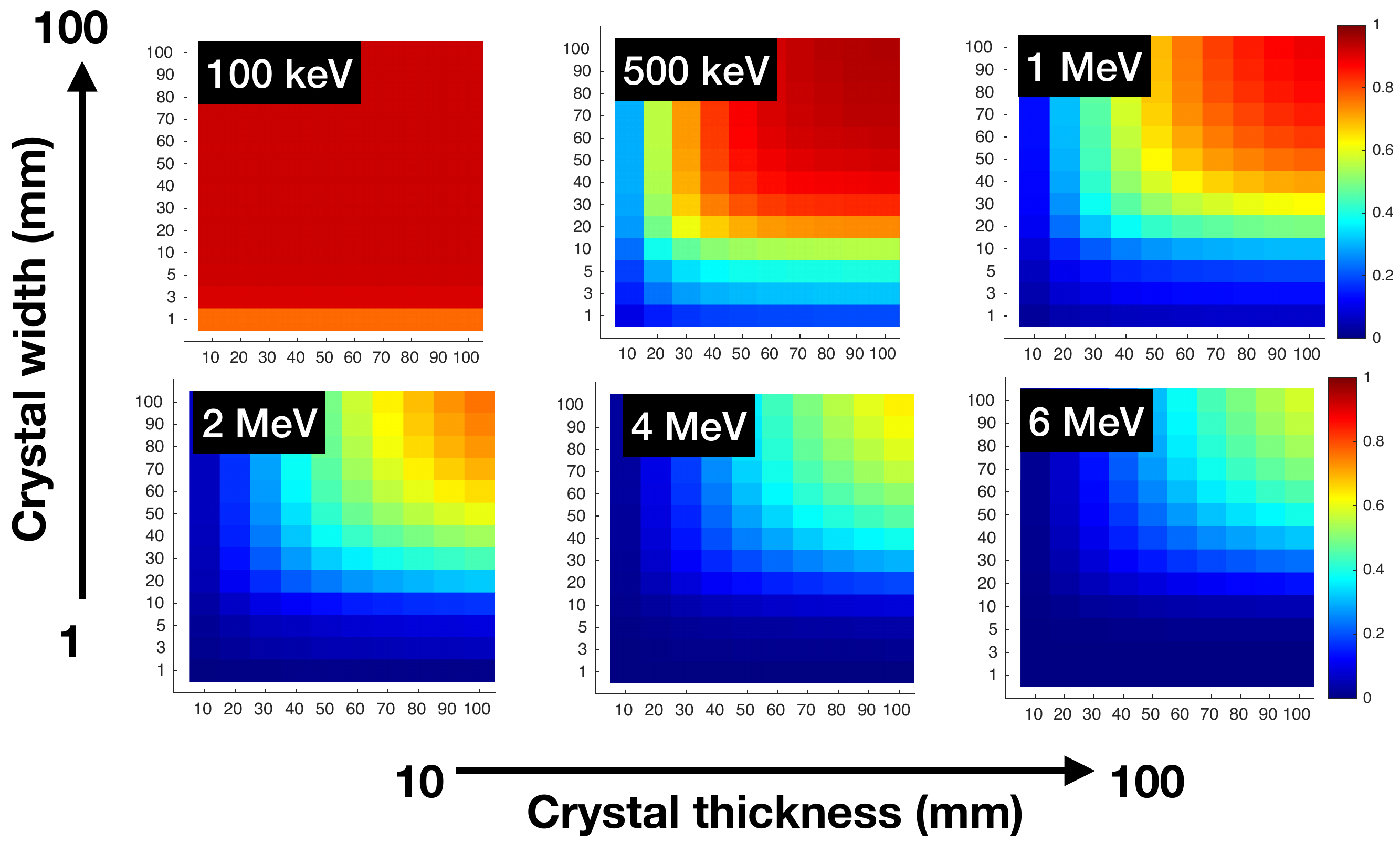}
	\caption{}
	\end{subfigure}
	\hfill
\caption{(a): Energy dependent interaction cross-sections corresponding to various process of photon interaction inside LYSO. Generated using XCOM tool \cite{XCOM2}. (b) Top: Photopeak efficiency plots for LYSO crystals as a function of thickness and width (equal values widths were employed for a square shaped section) shown for various input gamma energies.}
\label{fig_Intro_Crystalsize-Efficiency}
\end{figure}

Gamma photons interact with matter via three main processes Photoelectric, Compton and pair production. The mass attenuation coefficients for various gamma interaction processes in LYSO can be seen in figure \ref{fig_Intro_Crystalsize-Efficiency} (a) generated using XCOM photon cross section data base developed by \cite{XCOM1} , \cite{XCOM2}.  With increasing photon energy, the photoelectric conversion probability decreases. As Compton scattering leads to a partial deposition of energy in the crystal, multiple interactions in the same crystal can be fully recovered with a high probability if the crystal is sufficiently large. In order to understand the dependence of the percentage of events depositing the complete energy within a crystal on the crystal dimensions, we performed simulations on LYSO crystals of various sizes for different gamma energies as seen in  figure \ref{fig_Intro_Crystalsize-Efficiency} (b). The need for very large crystal sizes to recover a decent fraction of photons with energy higher than 5 MeV is clear. Photons under 1 MeV can recover around 50\% energy in the photopeak of crystals 5 cm wide and 4 cm thick. The 50\% photopeak efficiency for 5 cm wide crystals at 2  MeV  needs to be 6 cm long and for 5MeV gamma this thickness is  12cm. 10 cm wide crystals can achieve a similar efficiency at only 6 cm thickness. 

%
%

Localization of the incidence angle of high energy gamma is routinely achieved by the use of windows of high density materials as collimators which attenuate photons outside the window of acceptance. For a typical prompt gamma spectrum in the range of a few hundred keV to 6.14 MeV oxygen PG, $\sim10$ cm lead is required to achieve a complete attenuation.


Imaging proton induced gamma requires detectors with high efficiency and resolution. High-purity Germanium (HPGe)  detectors deliver the best energy resolution (0.3\% at 511 keV) but are significantly more expensive when compared to scintillator-photodetector systems. Other detectors such as Cadmium Zinc Telluride utilized in Compton imagers (\cite{Draeger2018}) have a good energy resolution but suffer from a lower efficiency and are expensive to manufacture in larger quantities with a reliable performance. 


Best choices  of inorganic crystals with regard to energy resolution are LaBr$_3$, CLLB, CLYC in the range of 1.5-3\% FWHM measured at 661 keV. In comparison,  LYSO has a poorer energy resolution of around 8\% FWHM at 661 keV. A high gamma attenuation makes LYSO the ideal choice of crystal as the detection efficiency is $\sim$ 1.5$\mathrm{\times}$ larger than LaBr$_3$ and other crystals with high energy resolution. Time resolution of scintillators can be useful in background rejection, up to 90\% was achieved by \cite{Verburg2013} by restricting the gamma events to a 1.5 ns window for a cyclotron running on an RF pulse with 9.4 ns period. A useful summary of the crystal properties of interest for gamma detection with fast timing is provided in \cite{Schaart2021} . Saint-gobain \cite{SaintGobain} specifies the crystal properties of a wide range of crystals along with a useful efficiency calculator \cite{SaintGobainEfficiencyCalculator}.

The choice of crystal eventually depends on the range of gamma energies of interest and the requirement at hand. For a proton therapy imaging environment, LYSO crystals can be a good choice for cases of low statistics such as PAG detection. As we will describe in  section \ref{section_KE}, these can be suitable for PG detection based purely on threshold cuts and not energy selection. For the fiducial marker approach  described in \ref{section_FM} the region of interest is within 1 MeV and LaBr$_3$ can be an good choice. The correct choice of crystal for 3DPG detection described in section \ref{section_3DPG} is tricky since efficiency and resolution are both important. We will describe our design based on LYSO due to a higher gamma attenuation.

\section{Approach A: Knife edge collimator with PET type  detector}
\label{section_KE}


The knife-edge collimator with PET aims to utilize a collimator that first creates a depth-varying intensity pattern and a PET type detector which recovers low energy conversion events (Compton scattered) that depend indirectly on the incident high energy gamma. The dependence of the measured peak on the target position will be shown as a means of verifying the proton range. The compactness of this detector and the application of PET detectors (smaller crystals) which are more widely available makes this approach practical for a wide-spread adoption. A schematic is shown in figure \ref{fig_KE_setup}. 


\subsection{Design and setup}
\begin{figure}[h] 
\begin{subfigure}[b]{0.49\linewidth}
\centering
\includegraphics[width=0.8\textwidth]{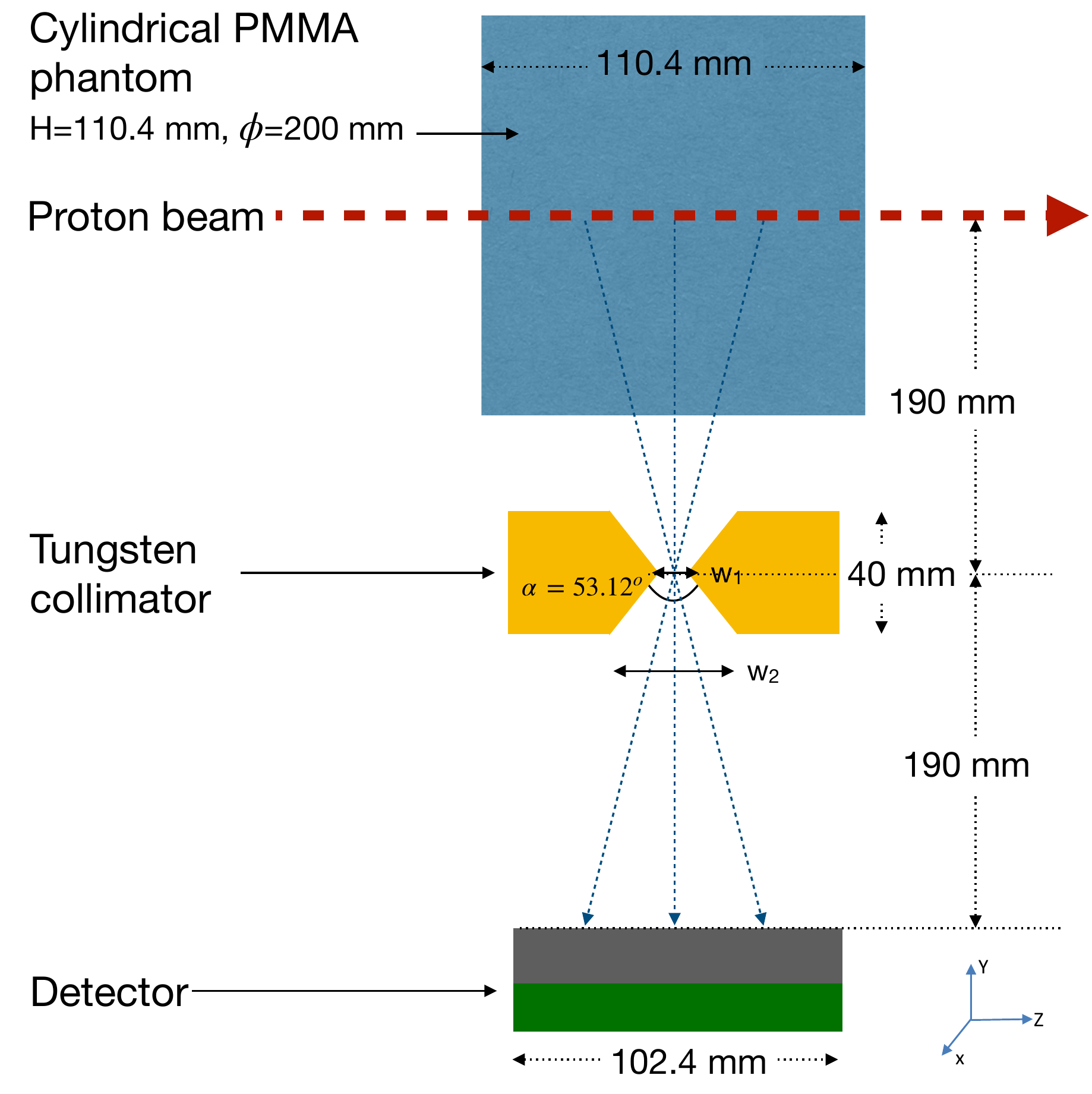}
\end{subfigure}
\hspace{0.02\linewidth}
\begin{subfigure}[b]{0.49\linewidth}
{\centering\includegraphics[scale=0.3]{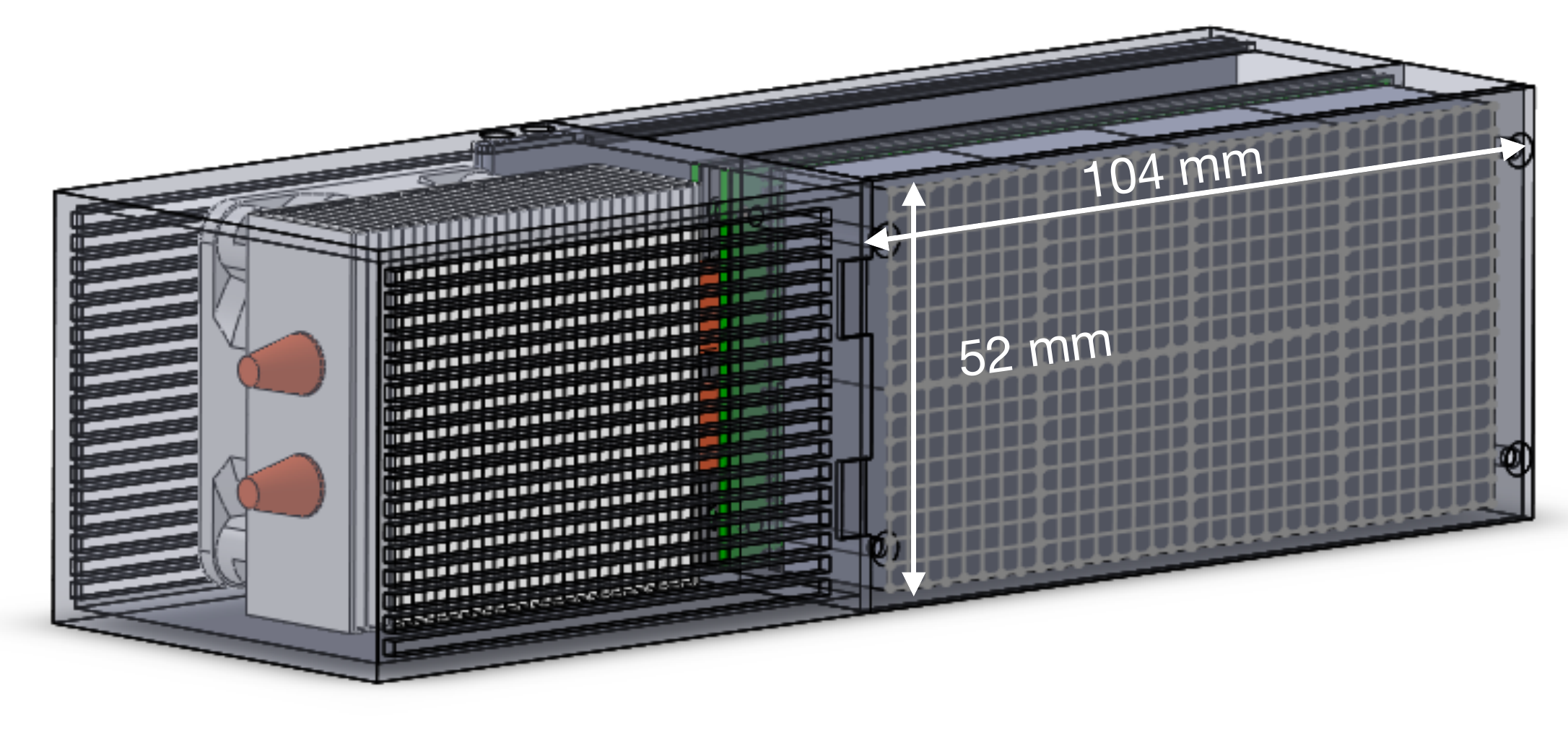}}
\end{subfigure}
\caption{Left: Schematic for setting up the knife-edge collimator  between a head-sized cylindrical phantom and a compact detector module. Right: Illustration of the ASPET palm sized module with 512 channels of 3mmx3mmx20mm LYSO crystals coupled one-to-one with SiPM arrays. }
\label{fig_KE_setup}
\end{figure}

In our simulated geometry, we irradiate a cylindrical phantom of PMMA (20 cm diameter 11.04 cm length) with the diameter comparable to a human head. An 8 cm thick tungsten knife-edge collimator is placed between the detector and the phantom with an area equal to the detector area. As seen in figure \ref{fig_KE_setup} (left) collimator inner  gap $\mathrm{w_1=15\ mm}$ and the outer width $\mathrm{w_2=35\ mm}$ with a resulting angle of $\mathrm{28.1^o}$.  The center of the  collimator is positioned at the expected Bragg peak position for 90 MeV proton beam. In a real treatment case, the Knife-Edge PET (KE-PET) system would need to be placed in a similar manner at the Bragg peak corresponding to the highest energy of the treatment plan. For the detector, we utilize the geometry of the ASPET as seen in \ref{fig_KE_setup} (right) described in \cite{Nemallapudi2021}. $\mathrm{3\times 3\times 20\ mm^3}$ LYSO crystals coupled directly to SiPMs with same area (3 mm $\times$ 3 mm) to form a single detector channel. A total of 512 channels (16 rows and 32 columns) are contained within one module and are readout by eight units of the 64 channel STIC asic \cite{stic} (developed in Univ. of Heidelberg) and the subsequent DAQ chain developed in-house supporting >2.5 MHz for a single module. In this work we model the detector on GATE and assume a uniform response and loss-less acquisition due to the low expected count rate.

When a gamma photon strikes any single channel, the entire energy is deposited during a photoelectric interaction, the probability of which decreases with the photon energy, and is $\sim 2\times$ lesser than Compton cross section at 511 keV.  The entire energy of a high energy gamma can also be recovered if multiple interactions occur within the same crystal leading to a full  recovery of the photon energy. This is the basis for using larger crystals in other PG detector designs. However, in our approach with the KE-PET, the counts of the partial energy deposited in each detector channel comprise the dominant part of the events. The counts retrieved in this manner are summed up along each vertical column of crystals.

\begin{figure}[h] 
\centering
	\begin{subfigure}[b]{0.48\textwidth} 
	\centering
	\includegraphics[width=\textwidth]{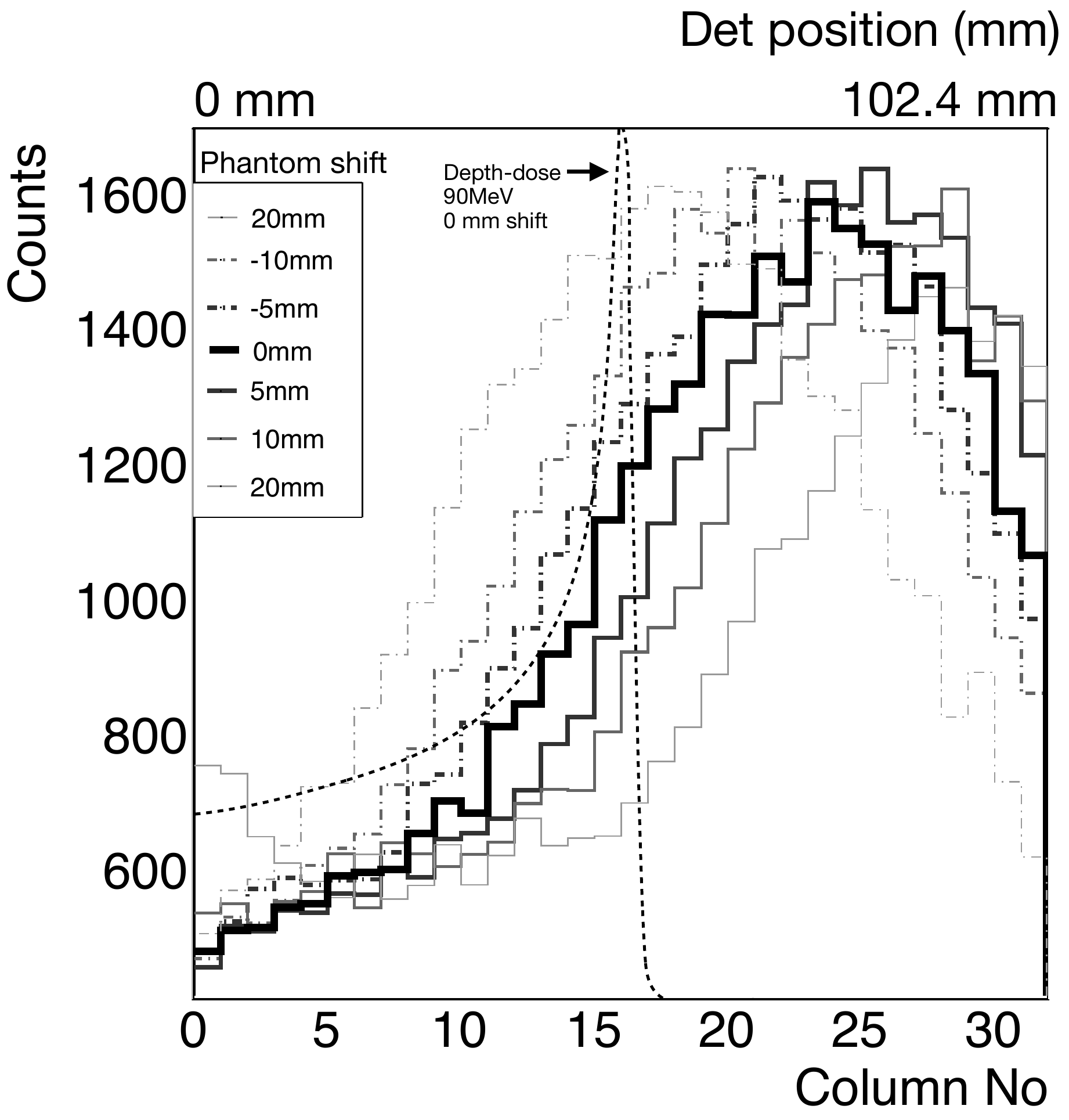}
	\caption{}
	\end{subfigure}
	\hfill
    \hspace{0.00\textwidth} 
	\begin{subfigure}[b]{0.46\textwidth} 
	\centering
	\includegraphics[width=\textwidth]{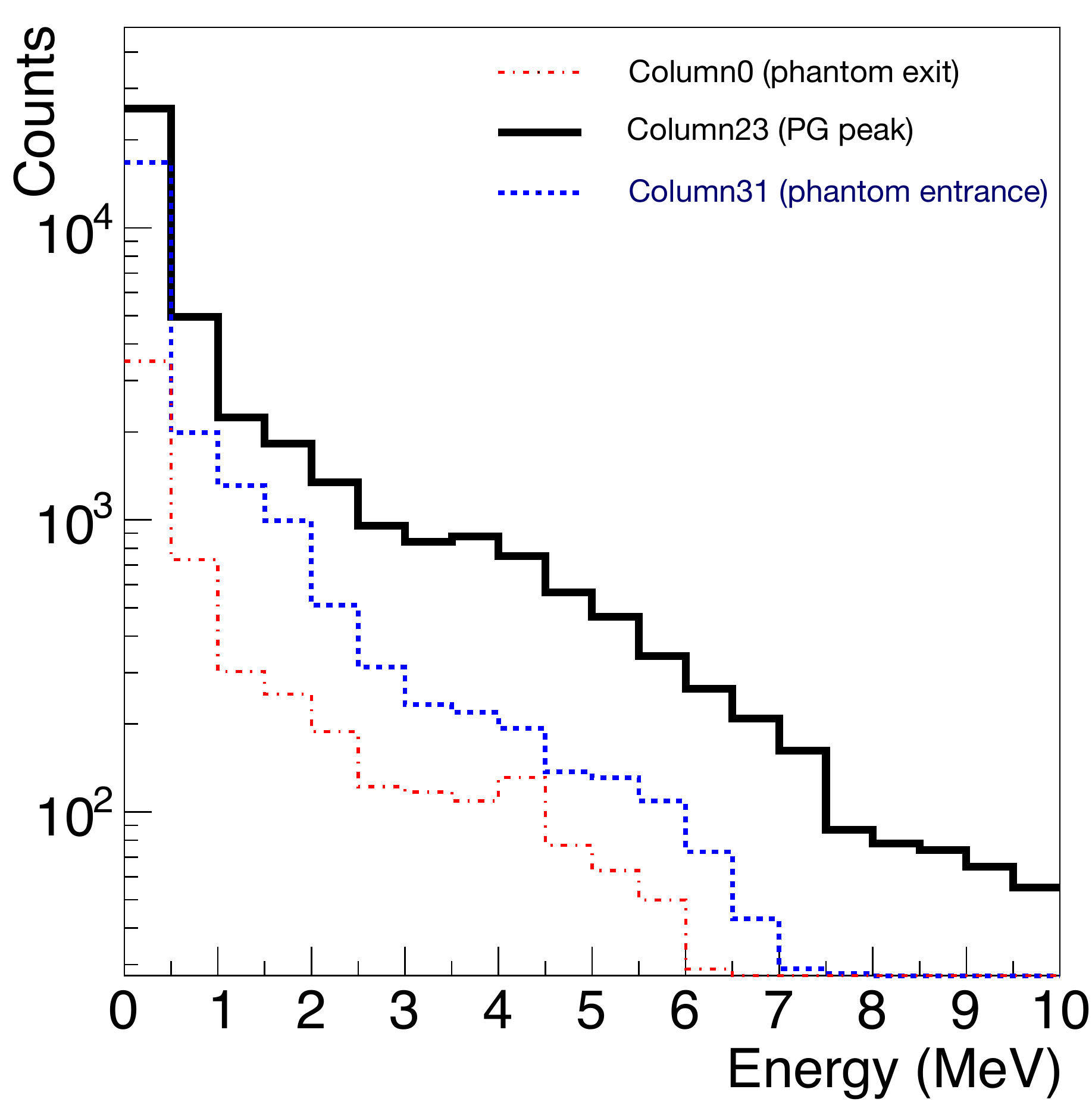}
	\caption{}
	\end{subfigure}
	\hfill
	\begin{subfigure}[b]{0.49\textwidth} 
	\centering
	\includegraphics[width=\textwidth]{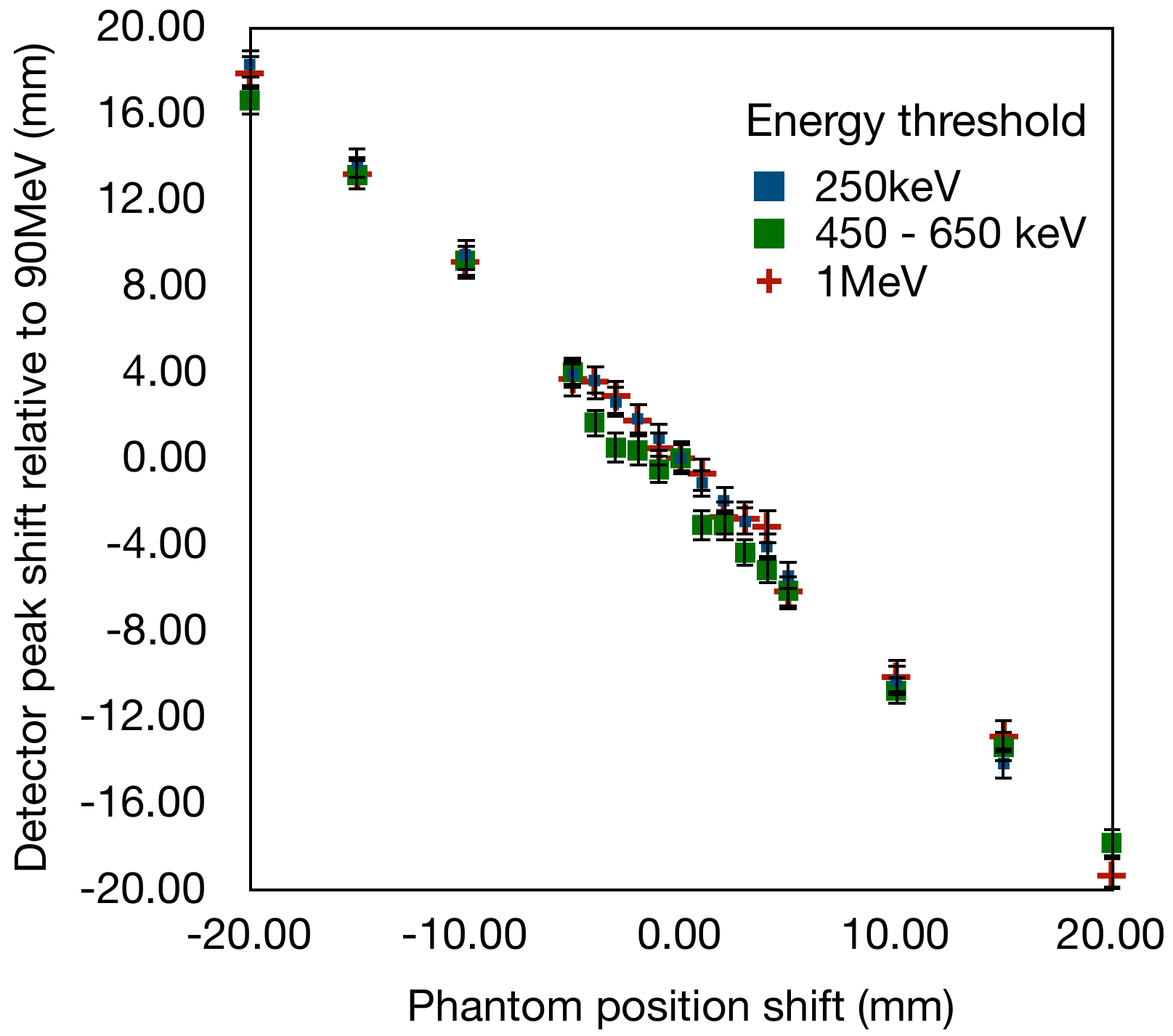}
	\caption{}
	\end{subfigure}\label{fig_3D_sector}
	\hfill
    \hspace{0.00\textwidth} 
	\begin{subfigure}[b]{0.49\textwidth} 
	\centering
	\includegraphics[width=\textwidth]{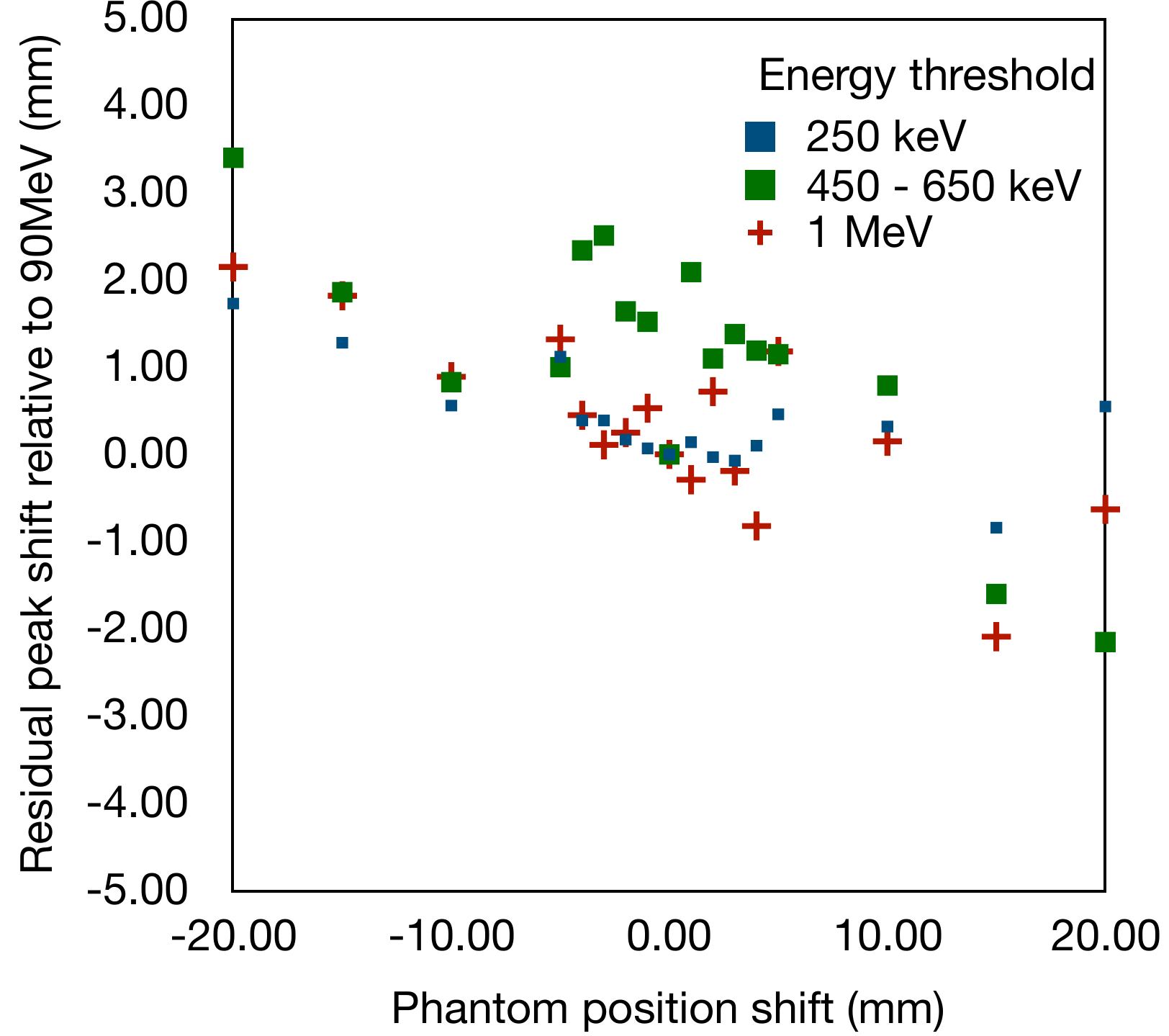}
	\caption{}
	\end{subfigure}\label{fig_3D_sector}
	\hfill
\caption{KEPET detector output simulated on Geant4 with 2 billion primary particles in accordance to the setup shown in figure \ref{fig_KE_setup}. (a) Integrated counts for each depth position recorded by the detector and plotted as a function of the crystal number/ position along the beam axis. Depth dose in the reference frame of the detector is shown as an eye guide for the nominal case of 0 mm range shift with a 90 MeV proton beam on 110.4 mm long PMMA. (b) Summed energy spectrum recorded in a detector column with 16 channels each  with $\mathrm{3\ mm\ \times 3\ mm\ \times\ 20\ mm}$ LYSO crystals. The effect of the PG from the target entrance, peak and the exit regions is shown in the corresponding locations of the detector. (c) Detector profile peak position for various positions of phantom shifts corrected for the nominal target position at  90 MeV plotted as a function of the phantom shift position for different energy threshold on the detector (d) Residual values for each shifted position with respect to the 45 degree slope value as a function of phantom position shift, shown for three different threshold energy values on the detector.}
\label{fig_KE_depth}
\end{figure}

\subsection{Phantom shifts with 90 MeV protons (simulation)}
We performed simulations of the KE-PET detection system with a 90 MeV proton beam irradiated on a cylindrical water phantom and recorded the detector information for 1 mm shifts in the phantom position from -20 mm to +20 mm. 

For each case, energy thresholds are set on energy spectrum (see figure \ref{fig_KE_depth} (a)) of each detection channel to obtain the counts. Counts from the detectors in the same depth are summed up. A 1D depth distribution of these counts is then plotted for each case of phantom-shift as seen in figure \ref{fig_KE_depth} (b) and the peak positions are obtained with a Gaussian fitting function. Twice the value of the statistical error representing a 95 \% confidence interval is reported as the error bar for the peak position. The peak positions recorded in this manner for various shifts of the phantom are corrected from the central value by subtracting for each position the value of the peak position at the central position where the collimator center coincides with the NIST range for 90 MeV protons at 55.6 mm. The corrected peak positions for various phantom shifts are shown in figure \ref{fig_KE_depth} (c). The three cases shown are when the energy threshold set on each detector channel is 250 keV, 511 keV and 1  MeV. For an input of $\mathrm{2\times 10^9}$ protons,  we can estimate the peak position to within  $\pm$0.7 mm with a 95\% confidence interval for the  region presented with the 1 MeV threshold setting. The deviation of the corrected peak positions noted by the detector from the expected phantom shift position is plotted in figure \ref{fig_KE_depth} (d). The range shifts measured by the detector are close to the real shifts in the phantom position  in the region of -5 mm to +5 mm. For higher values of shifts on either side, up to 4 mm deviation can be seen. We can account for the larger areas by performing a calibration of the systematic deviations of the peak positions at various proton energies.

The thickness of the collimator, the gap at the narrowest and widest regions of the KE collimator and the distance of the collimator from the beam and the detector are the major factors that affect the detected image for a given gamma energy at a given location along the beam axis. Placing the collimator closer to the phantom causes a higher magnification in the generated image.

In the current work we establish that a PET module can be conveniently adapted to develop a practical range-verification system. In order to further improve the positional resolution of the KE-PET, detector channels in coincidence can be tagged into clusters. The centroid of the cluster can be used to better localize the position. 

\section{Approach B: Fiducial marker emissions\label{FiducialMarker}} 
\label{section_FM}

Implantable fiducial markers can be exploited for their characteristic gamma lines resulting from their interaction with protons. In \cite{Freitas2021}, for e.g, the rectal balloon utilized in prostrate cancer treatment is used to identify the  prompt gamma emitted from Silicon. \cite{Bello2020} et al have previously monitored the gamma spectrum for  heavy ions with a 6 mm Ti slab inserted in water using a CeBr detection unit. Ti alloys are well suited for being introduced into the human body due to bio-compatability \cite{Elias2014}. A naturally forming oxide layer on the outer surface of Ti makes it corrosion resistant. In our work we performed position and energy scans with a Ti marker in combination with water to investigate the feasibility of using a thin titanium marker to identify the region of maximum proton dose deposition. Such a marker can be implanted either alongside a rectal balloon for prostrate cancer application, or in the mouth or  through esophagus to reach locations in close proximity to critical organs and potential treatment volumes. 

\subsection{$\mathrm{^{48}Ti}$ with water: Proton beam position scan (experiment) }
A beam test was conducted at a 30 MeV proton facility in INER using a HPGe detector (described in section \ref{section_ExperimentalSetup})  to identify the characteristic gamma energy lines and their relative expression in the gamma spectrum.  The dominant PG line expected from $\mathrm{^{48}Ti*->g.s}$ is the 984 keV gamma \cite{Kozlovsky2002}.

\begin{figure}[h] 
\centering
	\begin{subfigure}[b]{0.39\textwidth} 
	\centering
    \includegraphics[width=\textwidth]{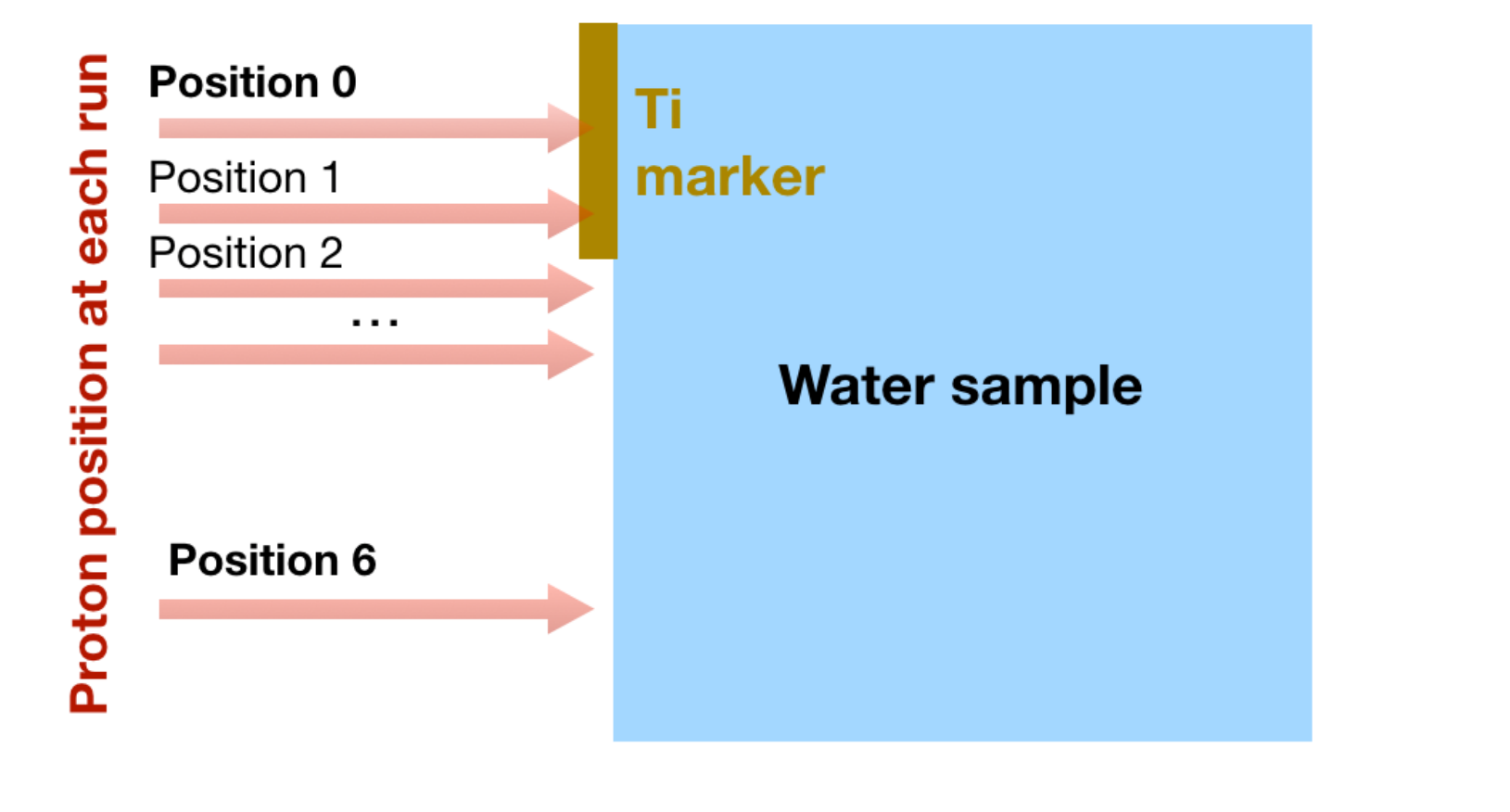}
	\caption{}
	\end{subfigure}
	\hfill
    \hspace{0.00\textwidth} 
	\begin{subfigure}[b]{0.59\textwidth} 
	\centering
	\includegraphics[width=\textwidth]{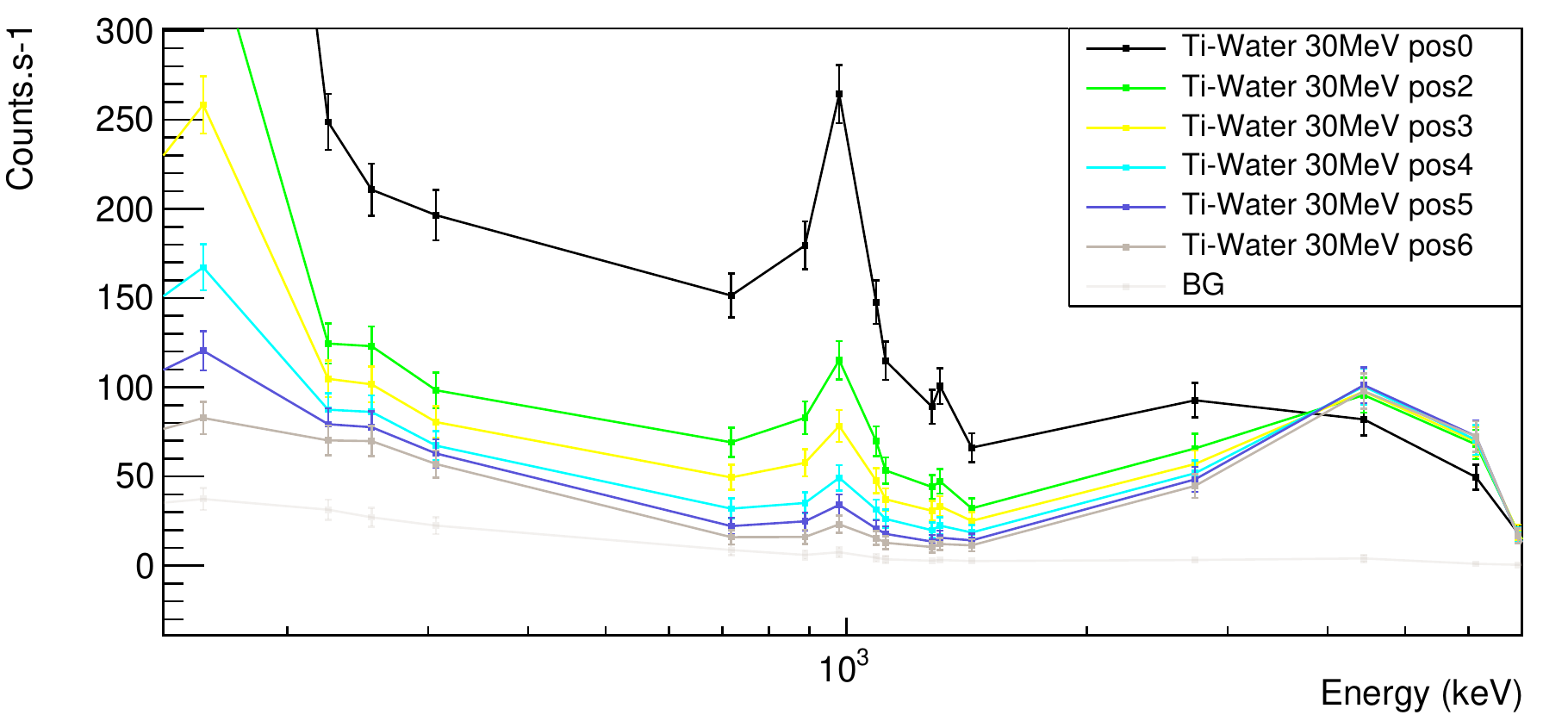}
	\caption{}
	\end{subfigure}
	\hfill
	\begin{subfigure}[b]{0.99\textwidth} 
	\centering
	\includegraphics[width=\textwidth]{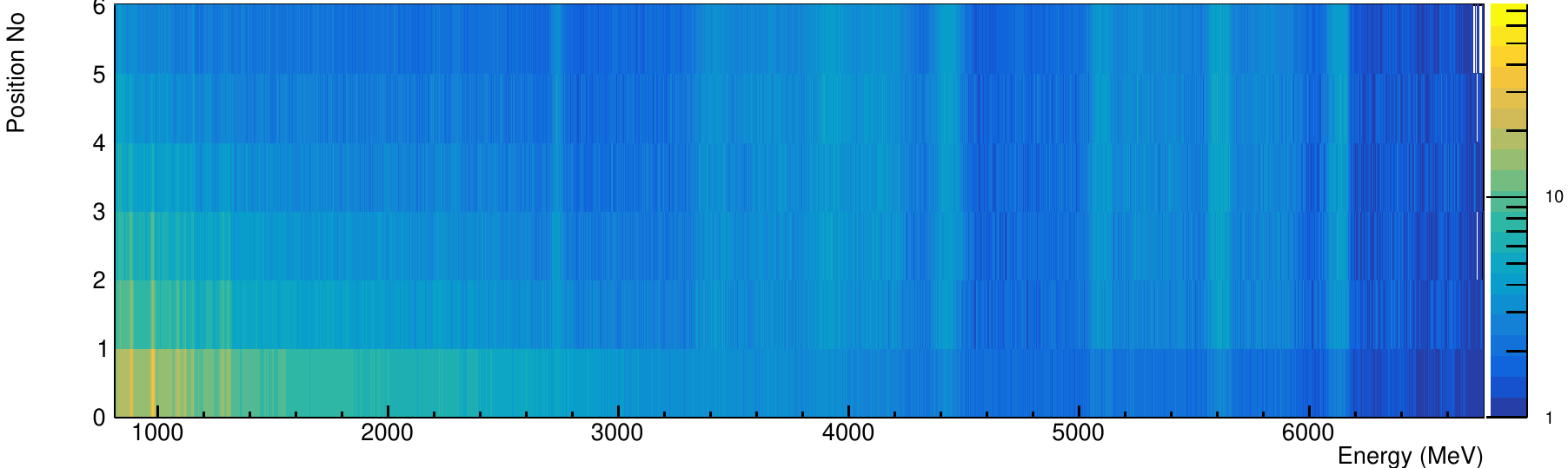}
	\caption{}
	\end{subfigure} 
	\hfill
\caption{(a) Schematic setup at INER proton facility delivering 30 MeV protons at different positions on a water phantom with an externally placed Ti  marker (1 mm thick). Beam was irradiated at different positions on the phantom surface. (b) Measured counts of the major gamma lines emitted by $\mathrm{^{48}Ti}$ at different positions of irradiation.  The region below 2 MeV is dominated by PG lines originating from Ti, while the region above 2 MeV mostly comprises PG from water. (c)Measured energy spectra at various positions of proton irradiation shown as a 2D intensity plot. }
\label{fig_FM_INER}
\end{figure}


The major lines of the Ti target are 158 keV, 308 keV, 984 keV along with minor lines at 880 keV, 1.3 MeV.

A titanium marker in the experimental test was subsequently placed along with a water phantom (water enclosed in a 0.1 mm thick PET bottle) to observe the Ti gamma emission  against a water backround by irradiating the protons  at different locations. As seen in the figure \ref{fig_FM_INER} (a), the major and minor gamma lines emitted by the $\mathrm{^{48}Ti}$  target can be discerned against the water background. When the proton beam is irradiated at positions close to the Ti target, the gamma lines produced by Ti are dominant in the spectrum. As the position of the beam irradiation is moved away from the Ti target, the PG lines emitted by water  are seen to be the major contributors while the intensities  of the Ti gamma lines reduce to zero. On \ref{fig_FM_INER} (b) the values plotted are the integrated counts in the energy spectrum about the region of interest of $\mathrm{\pm2\sigma}$ from each peak including the background contribution. In this figure a high level of contrast can be seen in the counts at 984 keV. The details of the detector are described inn section \ref{section_DetectorDesign}

Figure \ref{fig_FM_INER}  (c) describes the energy spectrum shown in a 2D histogram for gamma energy bins along X axis and the different positions (0-5) of irradiation along the y axis. The lines corresponding to $\mathrm{^{12}C}$ at 4.438 MeV and $\mathrm{^{16}O}$  at 6.129 MeV are clearly visible along with their single and double escape peaks. For position-0 which at which the protons are  directly incident on the Ti target, the 984 keV gamma line is prominent, while the gamma lines from water ($\mathrm{^{16}O^*->g.s:6.129\ MeV}$, $\mathrm{^{16}O(p,3p2n)^{12}C^*->g.s:4.438\ MeV}$) have a  lower intensity. The strong 511 keV line from PAG generation within the water phantom can be  noted in figure \ref{fig_FM_INER} (b). The main conclusion from the low energy measurement with Ti placed on top of a water phantom is that the major  gamma lines resulting from the Ti marker have the energies under 1 MeV and can be easily discerned from a water background.

\subsection{$\mathrm{^{48}Ti}$ inside water: Proton beam energy scan (experiment) }
This test was repeated at a higher energy in Chang Gung Memorial Hospital by probing the water phantom (8 cm $\mathrm{\times}$ 8 cm area and 30 cm depth, ) containing a 3 mm thick Ti target placed between 42 mm and 45 mm from the beam entrance, with various proton energies (70MeV - 90 MeV). 

\begin{figure}[h] 
\centering\includegraphics[scale=0.5]{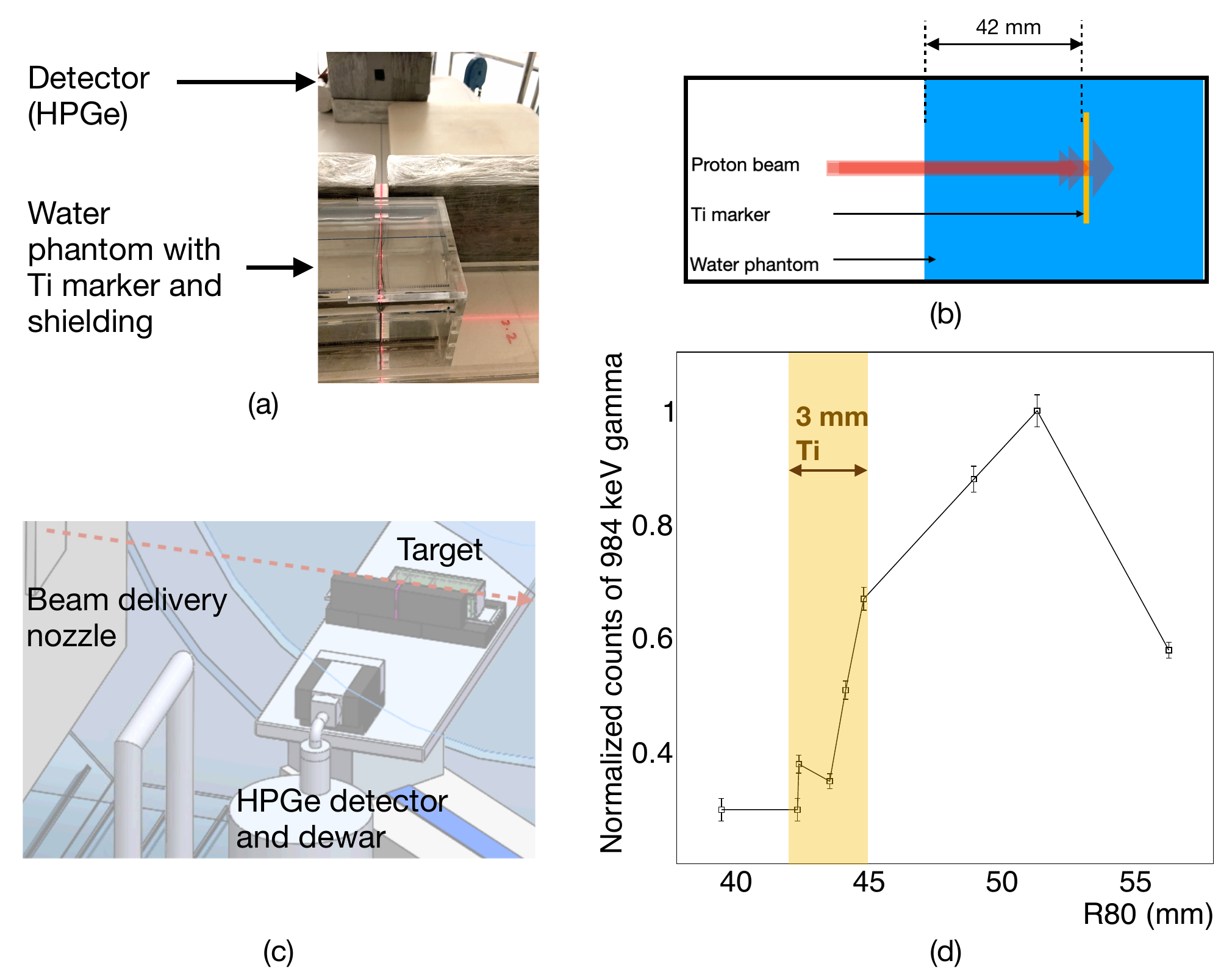}
\caption{(a), (b) and (c): Experimental setup at CGMH proton therapy facility shown in different views. Proton beam was irradiated on a water phantom with an inserted Ti  marker (3 mm thick) placed between 42 mm and 45 mm from the beam entrance. (d): The measured counts of 984 keV gamma line emitted by $\mathrm{^{48}Ti}$ as a function of the R80 depth (80\% of the peak dose on the distal side ). The gamma counts have been normalized to the incident beam current and irradiation time and the normalized values are plotted relative to the peak.}
\label{fig_FM_CGMH}
\end{figure}

The setup is shown in \ref{fig_FM_CGMH} (left) where the Ti target was placed inside a water phantom positioned on the treatment couch. 5 cm thick lead shields were placed adjacent to the water phantom facing the detector with a 2 cm window centered at the Ti target. The HPGe detector was positioned 46 cm away from the Ti marker shielded by lead bricks 5 cm thick and a circular entrance window of $\mathrm{2\ cm}$ diameter. The measurement was performed with different energies of proton beam for different runs. The beam currents and irradiaiton period were different for each proton energy - currents in the range of 0.27 nA to 0.42 nA, and irradiation period between 123 s and 196 s as shown in table \ref{table_FM_settings}. 

\begin{table}[]
\begin{center}
\caption{Experimental settings used in the test with proton irradiation on a water phantom with an inserted Ti marker. } \label{table_FM_settings}
\begin{tabular}{llllll} 
\hline
{\bf Beam   } & {\bf Beam   } & {\bf R80 } & {\bf Irradiation  } \\
{\bf  energy  } & {\bf  current  } & {\bf  position} & {\bf  time } \\
{\bf (MeV)  } &  {\bf (nA)  } & {\bf (mm)} & {\bf (s) } \\
\hline
70    & 0.267 & 39.4 &  133\\
74    & 0.297 & 42.3 &  123\\
74.8 & 0.299 & 42.3 &  164\\
76.4 & 0.301 & 43.5 &  196\\
78    & 0.383 & 44.1 &  190\\
80    & 0.389 & 44.8 &  162\\
84    & 0.400 & 48.9 &  151\\
86    & 0.406 & 51.3 &  128\\
90    & 0.418 & 56.3 &  183\\

\hline
\end {tabular}
\end {center}
\end {table}
\normalsize

The proton energy was then converted into a corresponding value of the R80 position which is the distal position with 80\% of the peak dose value. In the region of 960 keV to 1000 keV, a linear+Gaussian function is fitted to account for the background to obtain the counts of 984 keV gamma. The presence of the 984 keV gamma line as a function of the R80 value of the incident beam can be seen in figure \ref{fig_FM_CGMH} (right). The R80 position for the incident beams were estimated based on MC simulated depth doses for the test  beam setup of the water phantom with titanium marker for various input proton energies.  As seen in figure \ref{fig_FM_CGMH} d the counts 984 keV gamma originating from $\mathrm{^{48}Ti}$ increase from 30\% to 67\% of the maximum value when the proton range falls within the extent of the implanted marker. 

\subsection{Experimental measurements of PG production cross-sections}

\begin{figure}[h] 
\centering
	\begin{subfigure}[b]{0.49\textwidth} 
	\centering
	\includegraphics[width=\textwidth]{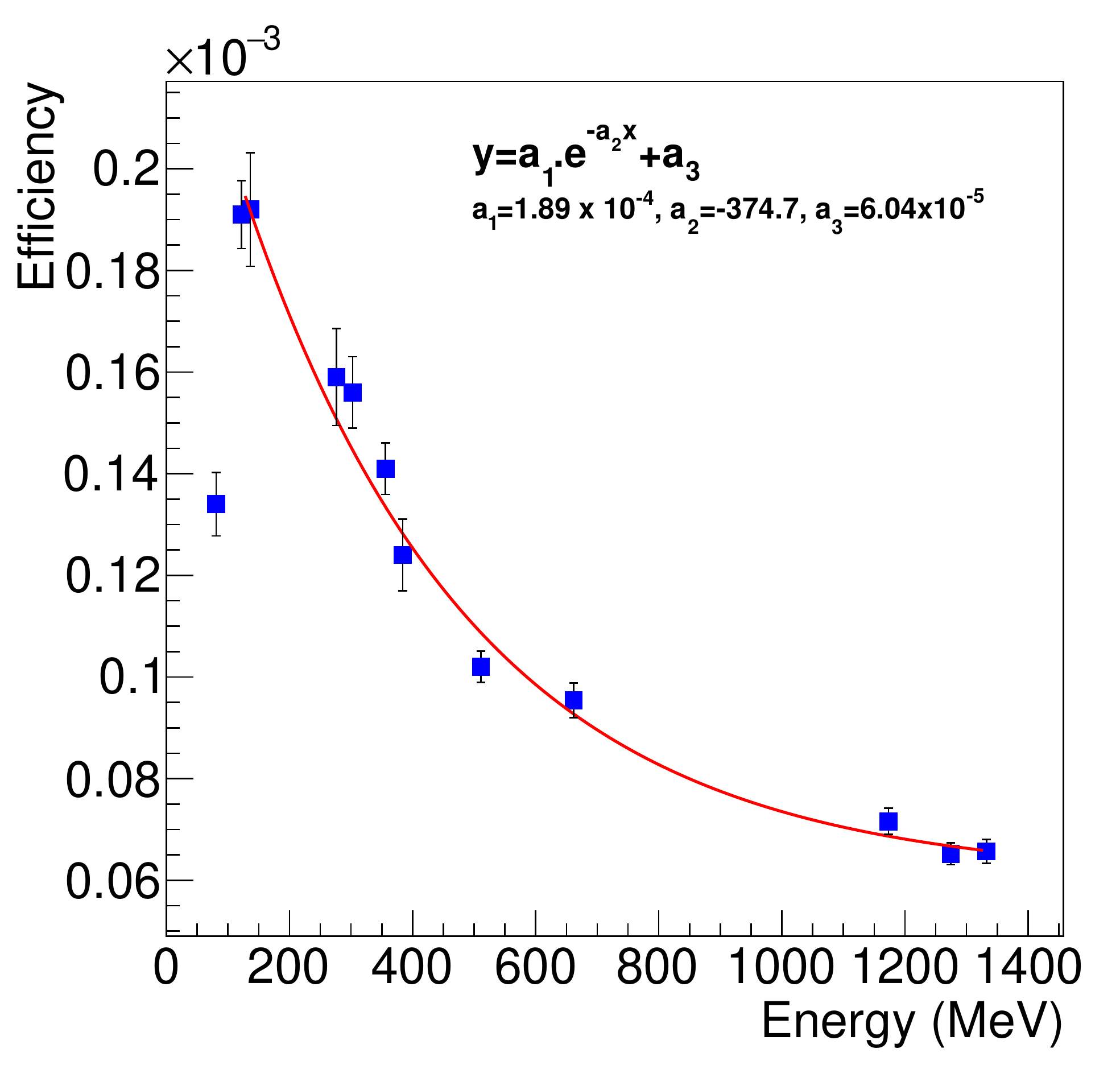}
	\caption{}
	\end{subfigure}
	\hfill
    \hspace{0.00\textwidth} 
	\begin{subfigure}[b]{0.49\textwidth} 
	\centering
	\includegraphics[width=\textwidth]{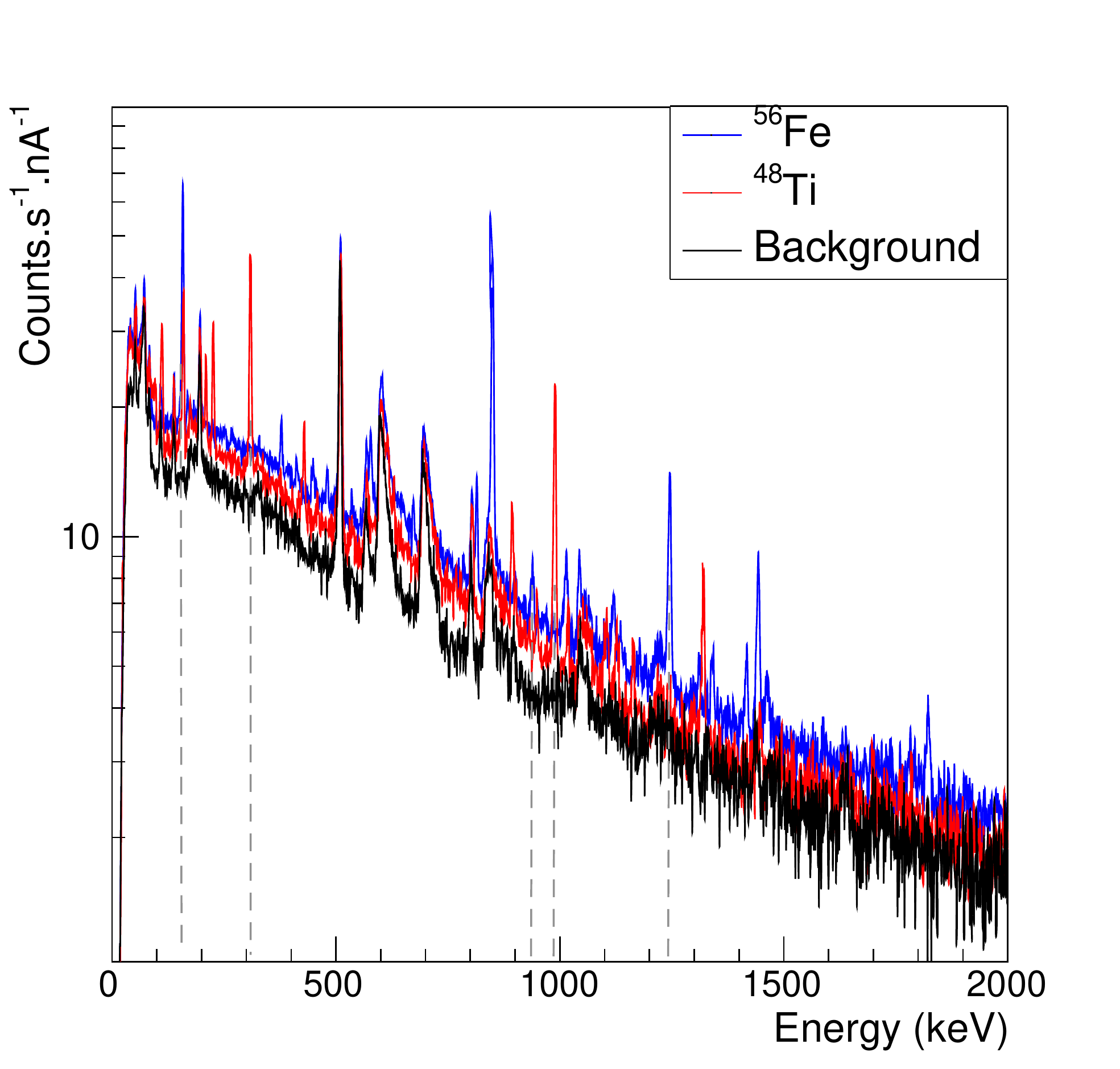}
	\caption{}
	\end{subfigure}
	\hfill
	\begin{subfigure}[b]{0.49\textwidth} 
	\centering
	\includegraphics[width=\textwidth]{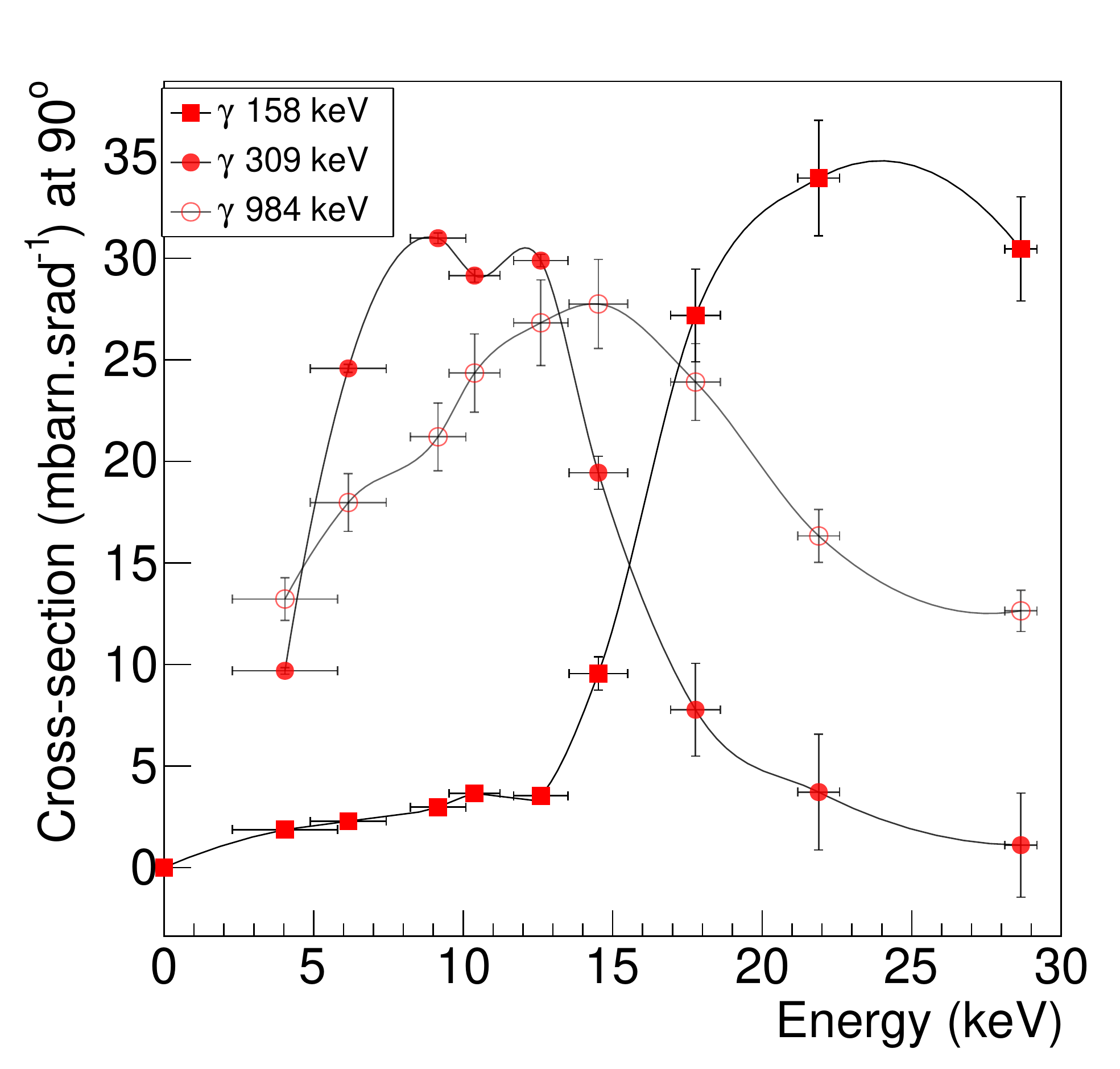}
	\caption{}
	\end{subfigure} 
	\hfill
	\begin{subfigure}[b]{0.49\textwidth} 
	\centering
	\includegraphics[width=\textwidth]{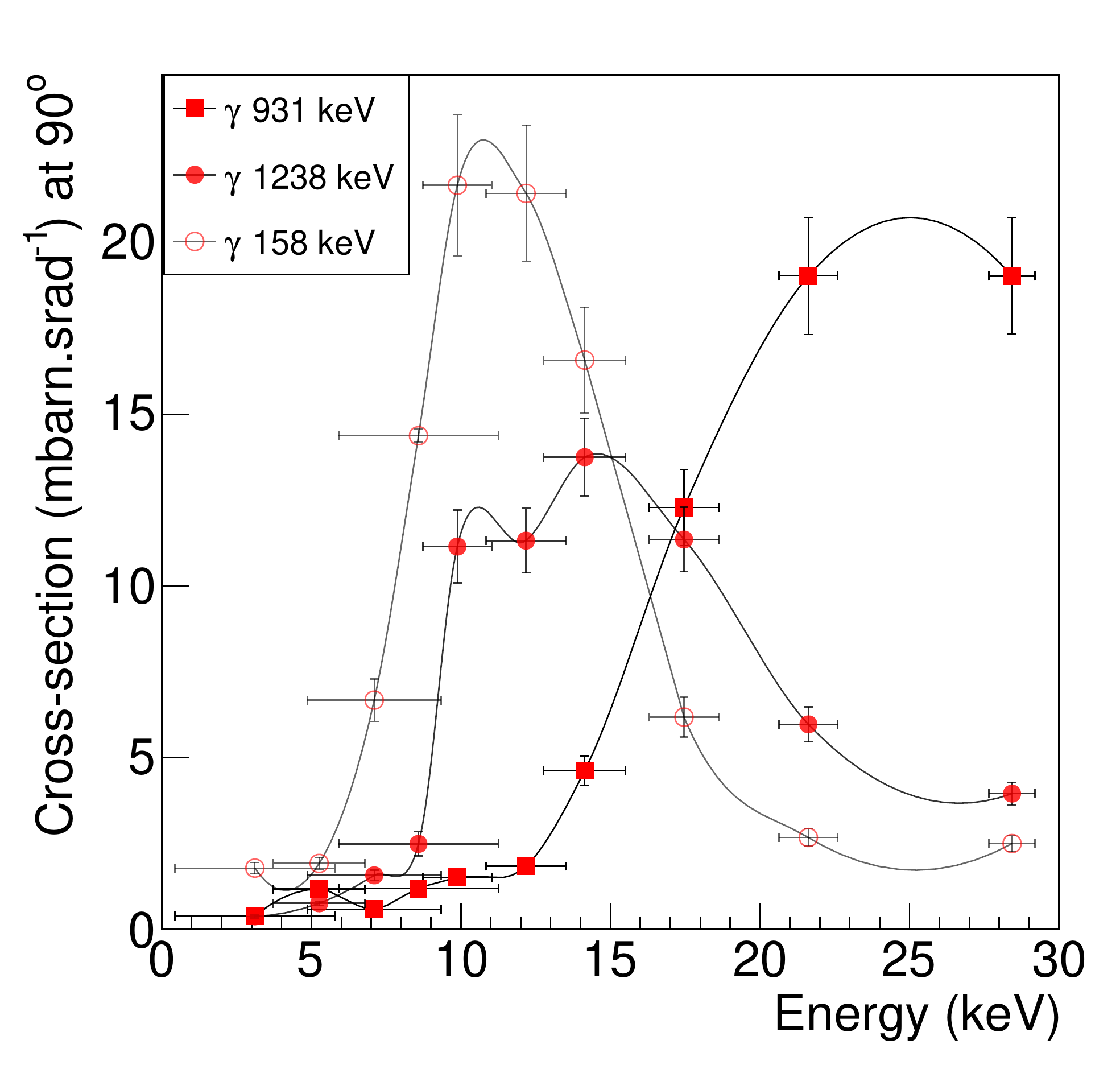}
	\caption{}
	\end{subfigure} 
	\hfill
\caption{Measurement of differential cross sections for prompt gamma emissions at a 90$^o$ angle. a) Efficiency of HPGe on standard sources $\mathrm{^{57}Co}$, $\mathrm{^{133}Ba}$, $\mathrm{^{22}Na}$, $\mathrm{^{137}Cs}$, $\mathrm{^{60}Co}$ for the same geometry. b) Energy spectrum of $\mathrm{^{48}Ti}$ and $\mathrm{^{56}Fe}$ PG emission for 14 MeV protons plotted along with the background (BG) in the absence of any target. c)Energy dependent cross sections of $\mathrm{^{48}Ti}$. d) Energy dependent cross sections of $\mathrm{^{56}Fe}$.}
\label{fig_FM_CS}
\end{figure}
In order to assess the possibility of applying the fiducial markers for range verification in a clinical setting it is necessary to estimate the expected counts which depend on the production cross-sections. We performed experimental measurements on two elements $\mathrm{^{48}Ti}$ since titanium can be good candidate for bio-compatible implants.

In an experimental test at INER, we measured the differential cross sections for the dominant PG lines at a $\mathrm{90^o}$ angle from  a  $\mathrm{^{48}Ti}$ target 0.1 mm thick and  50 mm $\times$ 50 mm area. 30 MeV protons were attenuated down to the desired proton energy before hitting the target. The prompt gamma emission was measured by a HPGe detector (described in section \ref{section_ExperimentalSetup}) placed 27.6 cm away from the target. The HPGe detection efficiency was measured against standard sources which matched with the region of interest as seen in figure \ref{fig_FM_CS} a.  The PG lines of interest were then fitted against the background spectrum measured without target and the counts were normalized against the beam current, irradiation time, the detection efficiency and the target thickness.

In order to calculate the energy dependent production cross-section of gamma we consider that the target area is larger than the beam, which is true for our measurement. 
We only consider prompt gamma in our calculations. The energy dependent cross section for each prompt gamma line can be estimated using equation \ref{eq_CS3}.
\begin{equation}\label{eq_CS3}
\mathrm{\sigma(E,90^o) = \frac{N_{det}(E)  A}{4\pi N_{protons} \eta_{det}  \rho N_{A}  dx},   }
\end{equation}
where E is the proton energy, $\mathrm{\sigma(E,90^o)}$ is the differential cross section at 90$^o$,  $\mathrm{N_{det}(E)}$ is the number of gamma detected, $\mathrm{\eta_{det} }$ is the detection efficiency, $\mathrm{\rho}$ is the material density, $\mathrm{N_{A}}$ is the Avogadro constant, $\mathrm{A}$ is the atomic mass, and $\mathrm{dx}$ is the thickness of the target.

The differential cross-sections measured at 90$^o$ can be seen in figure \ref{fig_FM_CS} c and d for titanium and steel targets. We estimated a peak values of $\mathrm{28\ mb.sr^{-1}}$  for 984 keV (15 MeV protons), $\mathrm{32\ mb.sr^{-1}}$  at 309 keV (10 MeV protons) and $\mathrm{34\ mb.sr^{-1}}$  at 158 keV (21.9 MeV protons)  impinging on a 0.1 mm thick, 50 mm $\times$ 50 mm area target of  $\mathrm{^{48}Ti}$.  In comparison, the 6.129 MeV PG from $\mathrm{^{16}O}$  and the 4.44 MeV PG from $\mathrm{^{12}C}$ from an oxygen target have differential cross sections values at $\mathrm{90^o}$ of around $\mathrm{11\ mb.sr^{-1}}$. The most reliable PG line from iron is the 1238 keV line resulting from an excited state of $\mathrm{^{56}Fe}$ and has a peak value of $\mathrm{14\ mb.sr^{-1}}$ as seen in figure \ref{fig_FM_CS} d. The low energy line at 158 keV with a peak at $\sim$10 MeV would be more sensitive to noise.

Based on the measured dependence and the expected counts, Ti marker based range verification can be found to be a practical approach. Stainless steel is another interesting candidate for this application.

In order to maximize the signal to noise ratio in a clinical application it would be necessary to employ larger crystals and utilize readout electronics with higher data acquisition rates. Anti-Compton shielding will be effective in suppressing the high energy gamma that leave the primary crystal, and also to reduce the neutron induced gamma which can come from larger angles with respect to the detector-target line of sight. There is sufficient scope to explore other options such as pixellated detectors or even miniaturized detectors that can be mounted closer to the fiducial marker.

\section{Approach C: 3D imaging of prompt gamma\label{3DPG}} 
\label{section_3DPG}

\begin{figure}[h] 
\centering 
\includegraphics[scale=0.35]{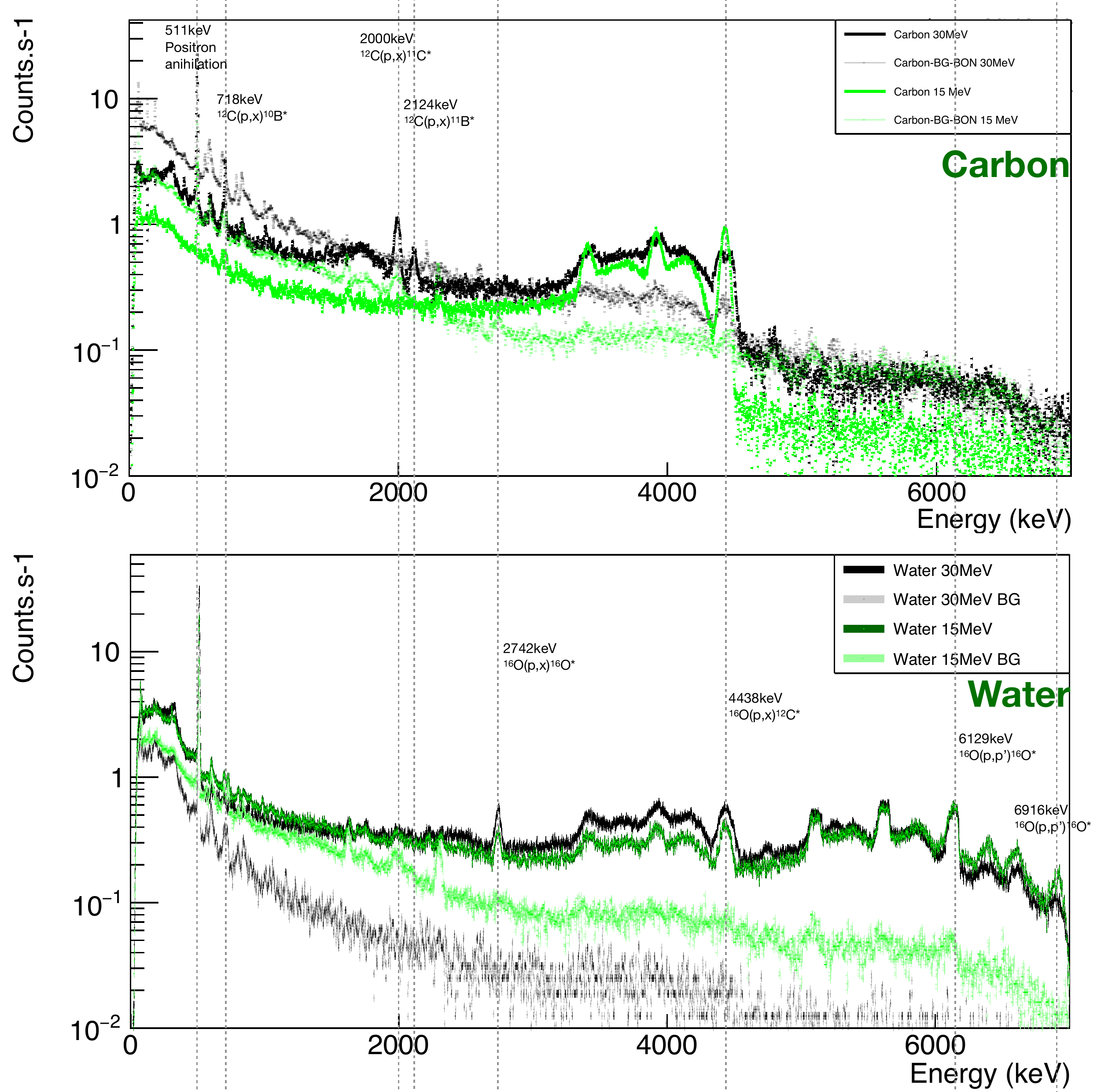}
\caption{Experiementally measured  prompt gamma spectrum produced by 30 MeV and 15 MeV protons interacting with carbon (top) and water (bottom). The background signal measured in the absence of any  target is also shown for comparison. The test was performed at INER.}
\label{fig_3DPG_Espectrum_Water-C}
\end{figure}

The prompt gamma emission is dependent on the target material and the energy of the interacting protons. Figure \ref{fig_3DPG_Espectrum_Water-C} shows various gamma lines emitted by carbon and water targets for two different proton energies (15 MeV and 30 MeV) in the region of highest cross sections for prompt gamma emission. The dominant gamma lines shown in this figure are most relevant for imaging a human patient comprising mostly of O and C. The expression of individual gamma lines changes along the target depth and results in a similarly evolving spectra as a function of proton energy. With a good energy resolution and a high spatial resolution individual energy lines can be tracked. The final application of this detection approach will be to utilize the cross section information to map the measured gamma to an inherent material distribution. We propose a detection scheme that can facilitate this.

In order to detect the proton-induced prompt gamma, a high efficiency detector with a large volume and a good resolution is required. Large detector volume enables the maximum recovery of the incident gamma energy. Since Compton scattering is dominant in the 1-5 MeV  region, recovery of the full energy of the incident photon to identify it requires a larger crystal volume. Localizing the position requires the usage of collimators that restrict the photons to a small region. Although using bulky collimators can provide better localization of the gamma incident angle, the detection efficiency is lost in the process. The design presented in this paper is an optimization between a high degree of localization and an acceptable detection efficiency. 

\subsection{Design and setup}\label{subsection_3DPG_design}

\begin{figure}[h]  
\centering
	\begin{subfigure}[b]{0.49\textwidth} 
	\centering
	\includegraphics[width=\textwidth]{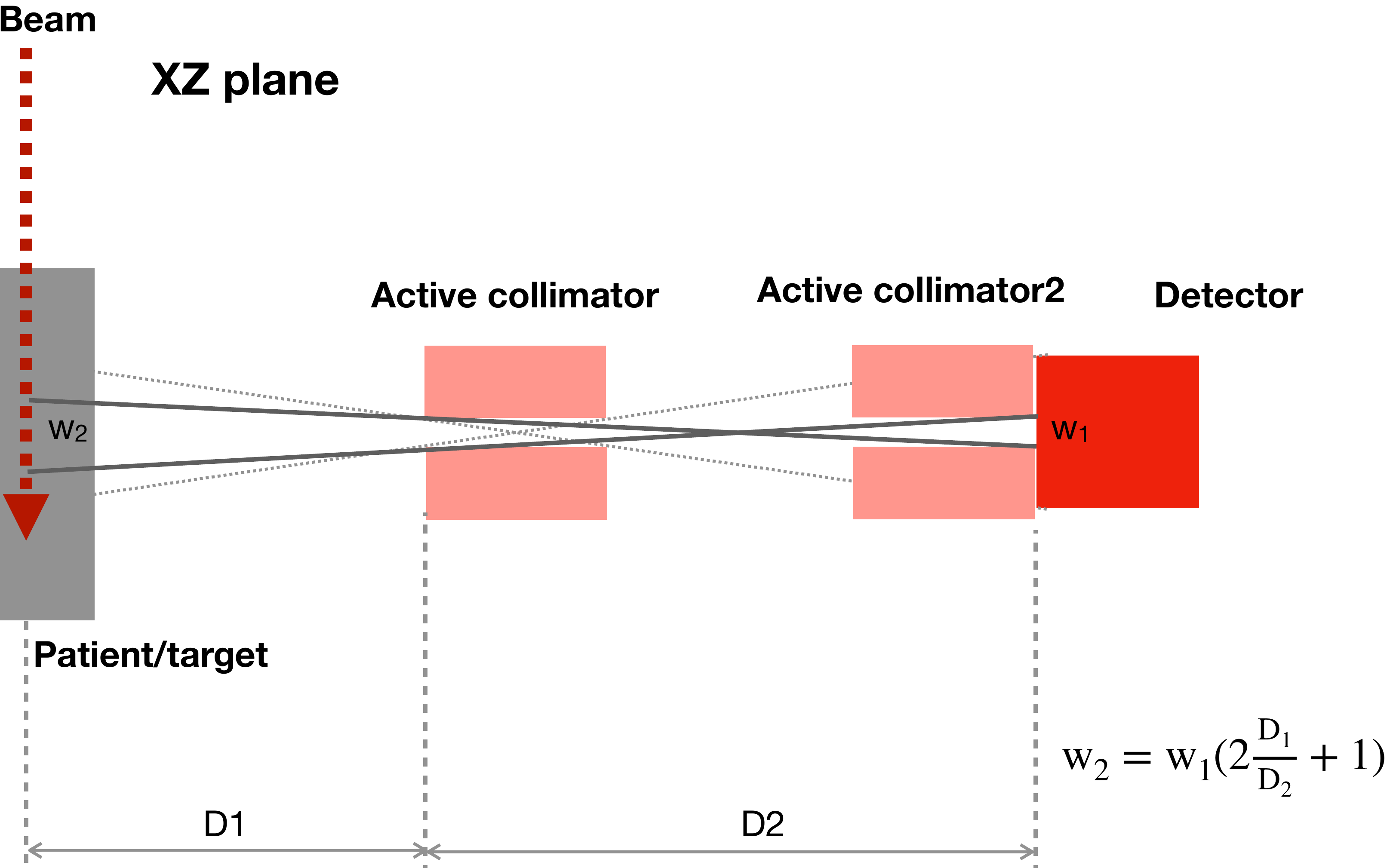}
	\caption{}
	\end{subfigure}
	\hfill
    \hspace{0.00\textwidth} 
	\begin{subfigure}[b]{0.4\textwidth} 
	\centering
	\includegraphics[width=\textwidth]{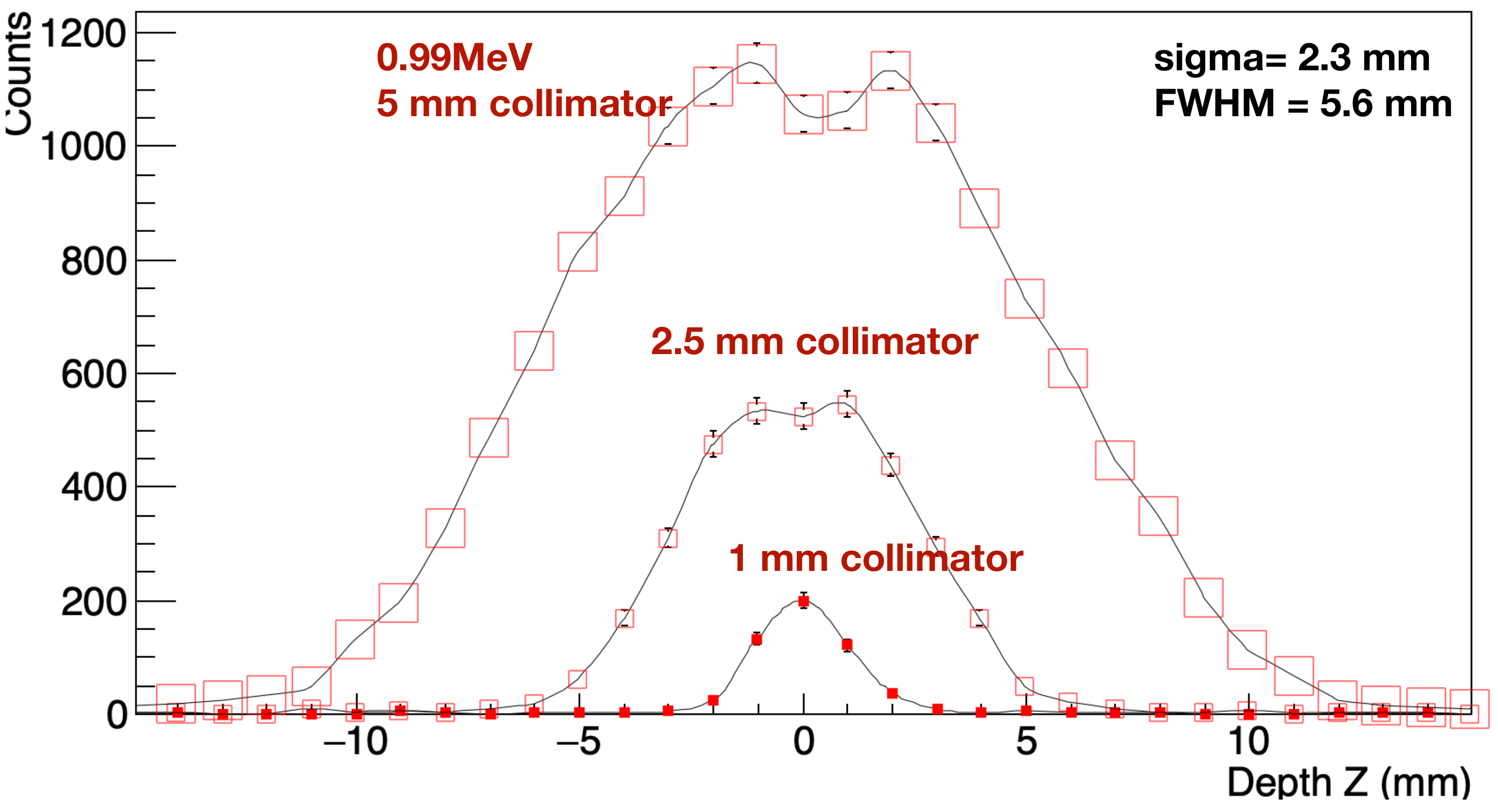}
	\caption{}
	\end{subfigure}
	\hfill
	\begin{subfigure}[b]{0.49\textwidth} 
	\centering
	\includegraphics[width=\textwidth]{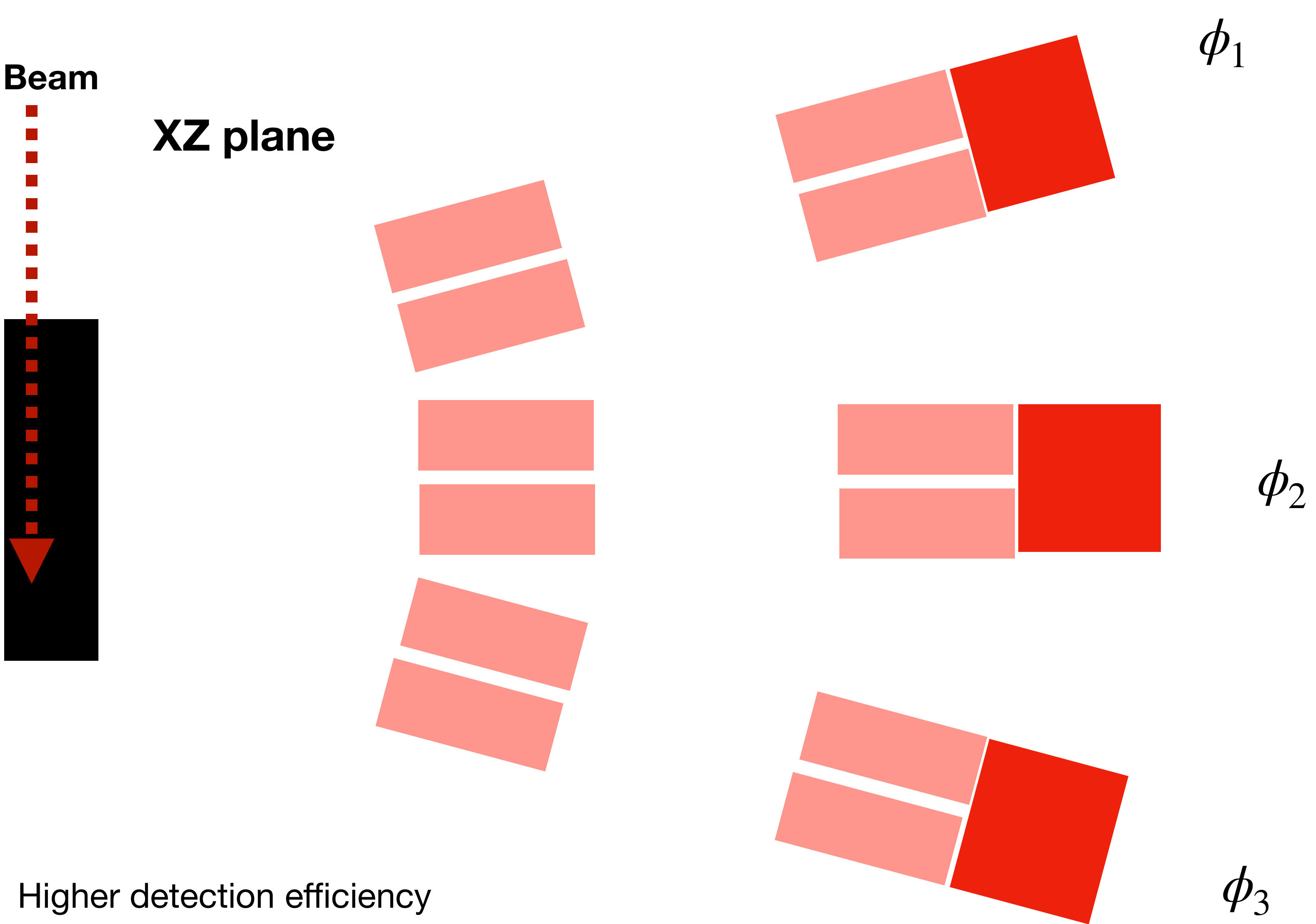}
	\caption{}
	\end{subfigure}\label{fig_3D_sector}
	\hfill
    \hspace{0.00\textwidth} 
	\begin{subfigure}[b]{0.49\textwidth} 
	\centering
	\includegraphics[width=\textwidth]{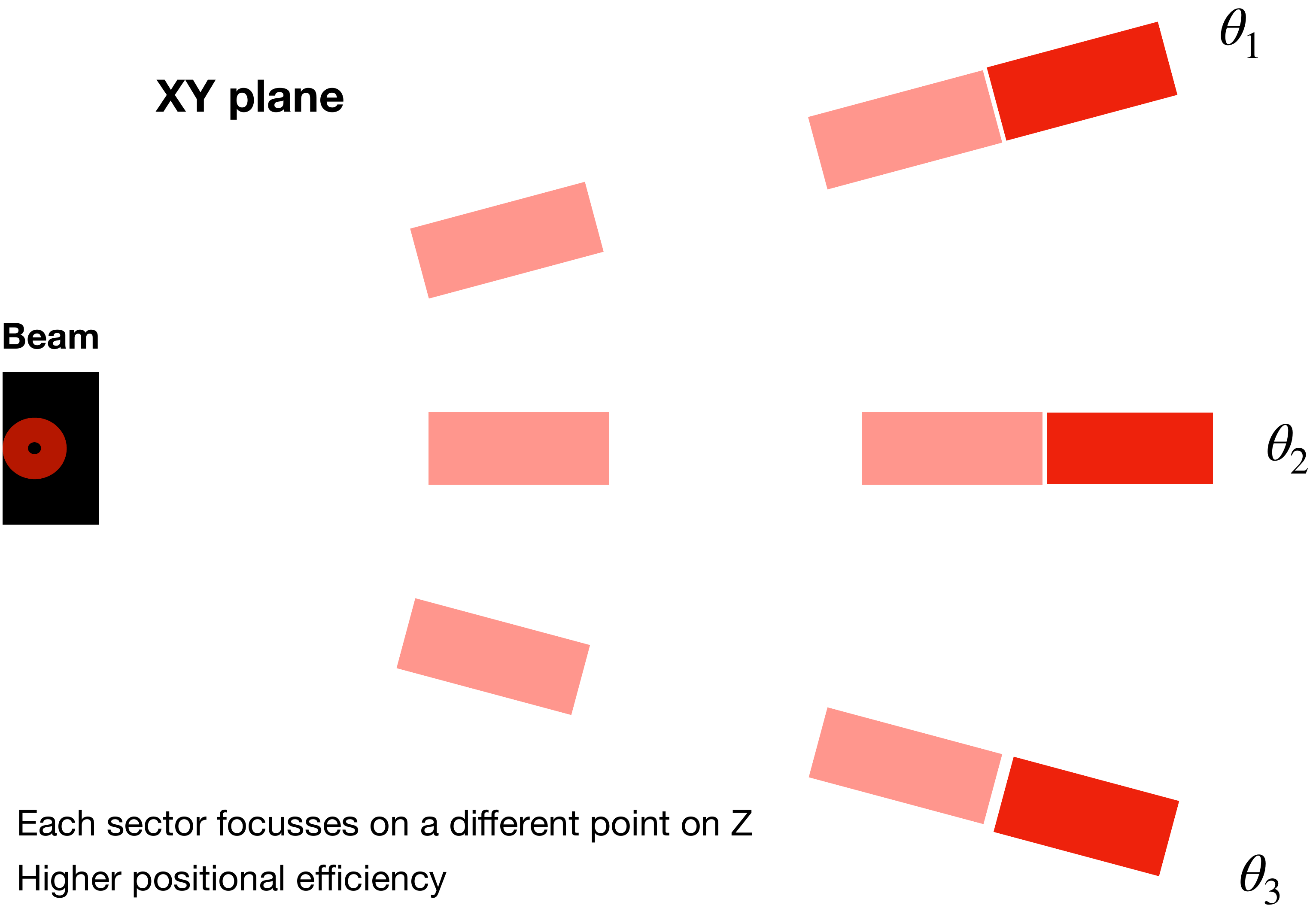}
	\caption{}
	\end{subfigure}\label{fig_3D_sector}
	\hfill
\caption{(a) Design of a single module with two stages of collimators. (b) Positional resolution of the located gamma source as a function of collimator width simulated with a 990 keV photon source. (c) Each sector made of multiple modules arranged along the same plane with $\phi$ symmetry about the point of interest. (d) Multiple sectors are placed about the point of interest at different $\theta$ angles.}
\label{fig_3D_singlemodule}
\end{figure}

To achieve  this a single module consisting of two stages  of collimators is presented as shown. A two stage collimation achieves better localization than a single stage as illustrated schematically  in figure \ref{fig_3D_singlemodule} (a). In order to image a patient the collimator needs to be sufficiently far to accommodate the patient size. For the presented work we chose the dimensions $\mathrm{D_1=30\ cm}$ and $\mathrm{D_2=33\ cm}$. The collimator gap is $\mathrm{w_1}$ and the region of interest within the target that is projected towards the detector is represented by $\mathrm{w_2}$. In order to restrict $\mathrm{w_2}$ to a smaller region, the ratio $\mathrm{\frac{D_1}{D_2}}$ needs to be minimized along with $\mathrm{w_1}$.
A larger value of $\mathrm{D_1+D_2}$ in turn leads to a lower detector efficiency. 

In our chosen design, the crystal for each module is a LYSO with $\mathrm{7\ cm\times 4\ cm \times 10\ cm}$ where the longest dimension 10 cm is in the axial direction, 7 cm is the crystal thickness along the radial direction and 4 cm in the theta direction. The choice of these dimensions is a compromise between detection efficiency for a single module, the ability to combine several sectors of detector, and cost considerations.  

The effect of collimator width has been studied for different collimator gaps that can achieve a millimeter level collimation. As seen in figure  \ref{fig_3D_singlemodule} (b) , the best position resolution was achieved for a 1 mm gap collimator at the cost of detection efficiency. The position resolution reported is the FWHM value of the distribution shown in figure  \ref{fig_3D_singlemodule} (b) which is the response function of the detector when a point source is moved along the axis of interest.  We have verified the effect of collimation experimentally using a NaI detector where two stage collimator of 5 mm gap yields a resolution of 9.7 mm FWHM for 511 keV gamma and 10.6 mm FWHM for 1.27 MeV gamma from a $\mathrm{^{22}Na}$ source. A single stage collimator in comparison provides a FWHM positional resolution of 19 mm for 511 keV gamma and 24.6 mm for the 1.27 MeV gamma.   The incident position uncertainty can be calculated as a function of the collimator gap and the distance between the collimators. Figure \ref{fig_3D_singlemodule} (b) shows the spatial resolution of the single module for various collimator gaps.

In order to increase the detection efficiency, multiple modules are added by shifting the original module along the azimuthal plane. This results in a planar sector of detectors focusing on the same point. This is shown in figure \ref{fig_3D_singlemodule} (c). In order to identify the depth distribution of the prompt gamma we need to identify several points with a similar level of detection efficiency. 
Multiple sectors are arranged with equal angle between consecutive sectors about the beam axis ($\mathrm{\theta}$ direction) as seen in  \ref{fig_3D_singlemodule} (d). A linear offset is then introduced for each sector with respect to the previous sector along the beam axis. Since each sector is designed to observe a specific region of the beam axis, multiple sectors with a linear offset will cover a larger range of positions along the beam.

\paragraph{Modes of operation:}
The sectors can be mounted on linear stages to position each sector to be sensitive to a specific position along the beam axis. As the number of sectors is limited, the detector can operate in three modes - a)  low efficiency, low resolution and high dynamic range or b) low efficiency high resolution and low dynamic range or c) high efficiency on a single location. 
In the first mode, the sectors are spaced with a larger gap along the z axis, this can scan over a larger beam range. In the second mode,  the sectors can be spaced with a smaller gap along the z axis and can observe a smaller region with a higher resolution.
In the third mode, all the sectors can be aligned to observe a single point along the beam axis. The flexibility of this design allows a seamless switching between different modes if a real-time imaging is preferred. 
Table  \ref{table_3D_detectionefficiency} lists the obtained resolution and photo-peak detection efficiency per each sector of the detector for different collimator gaps at two different energies of interest. The photo-peak efficiency scales linearly when more sectors are tuned-in to observe the same point. 
The detector design presented here is essentially a 1D detector capable of 3D imaging when run in tandem with a scanning pencil beam with the prior information on the location of the pencil beam.

\begin{table}[]
\begin{center}
\caption{A summary table of the spatial resolution and the region of interest (ROI) for localizing a point source on the axis of interest for three different collimator gaps. The photo-peak detection efficiency values for low energy and high energy gamma are shown. The entire setup was simulated on GATE/Geant4} \label{table_3D_detectionefficiency}
\begin{tabular}{llllll} 
\hline
{\bf Collimator  }  & {\bf ROI} & {\bf Positional } & {\bf Efficiency} & {\bf Efficiency} \\
{\bf gap w$_1$}  & {\bf  w$_2$ } & {\bf resolution} &  {\bf (sector$^{-1}$)}  &  {\bf (sector$^{-1}$)}  \\
{\bf mm}  & {\bf mm} & {\bf mm (FWHM)} & (at 990keV)  & (at 6.13MeV)    \\
\hline
1 & 2.8 & 2.6  &   $\mathrm{8.3\times 10^{-6}}$  &   $\mathrm{5.4\times 10^{-6}}$ \\
2.5 & 7.0 & 5.6  &   $\mathrm{2.2\times 10^{-5}}$  &   $\mathrm{1.1\times 10^{-5}}$ \\
5 & 14.1 & 11.7  &   $\mathrm{4.4\times 10^{-4}}$   &   $\mathrm{3.1\times 10^{-4}}$\\
\hline
\end {tabular}
\end {center}
\end {table}
\normalsize

\subsection{1D detection of proton induced PG (simulation)}
A detector model with 8 sectors covering 120 degrees with 3 modules per sector was simulated on GATE. The module dimensions were as described in section \ref{subsection_3DPG_design} with a collimator gap $\mathrm{w_1=2.5\ mm}$. Consecutive sectors are separated by 15 degrees about the beam axis, and linearly spaced along the beam axis by 2.5 mm. All the eight sectors put together cover a region of interest 17.5 mm along the beam axis in a continuous arrangement, or  in discrete arrangement if a larger region of interest is desired. 

\begin{figure}[h]  
\begin{minipage}[h]{0.49\linewidth}
{\centering\includegraphics[scale=0.2]{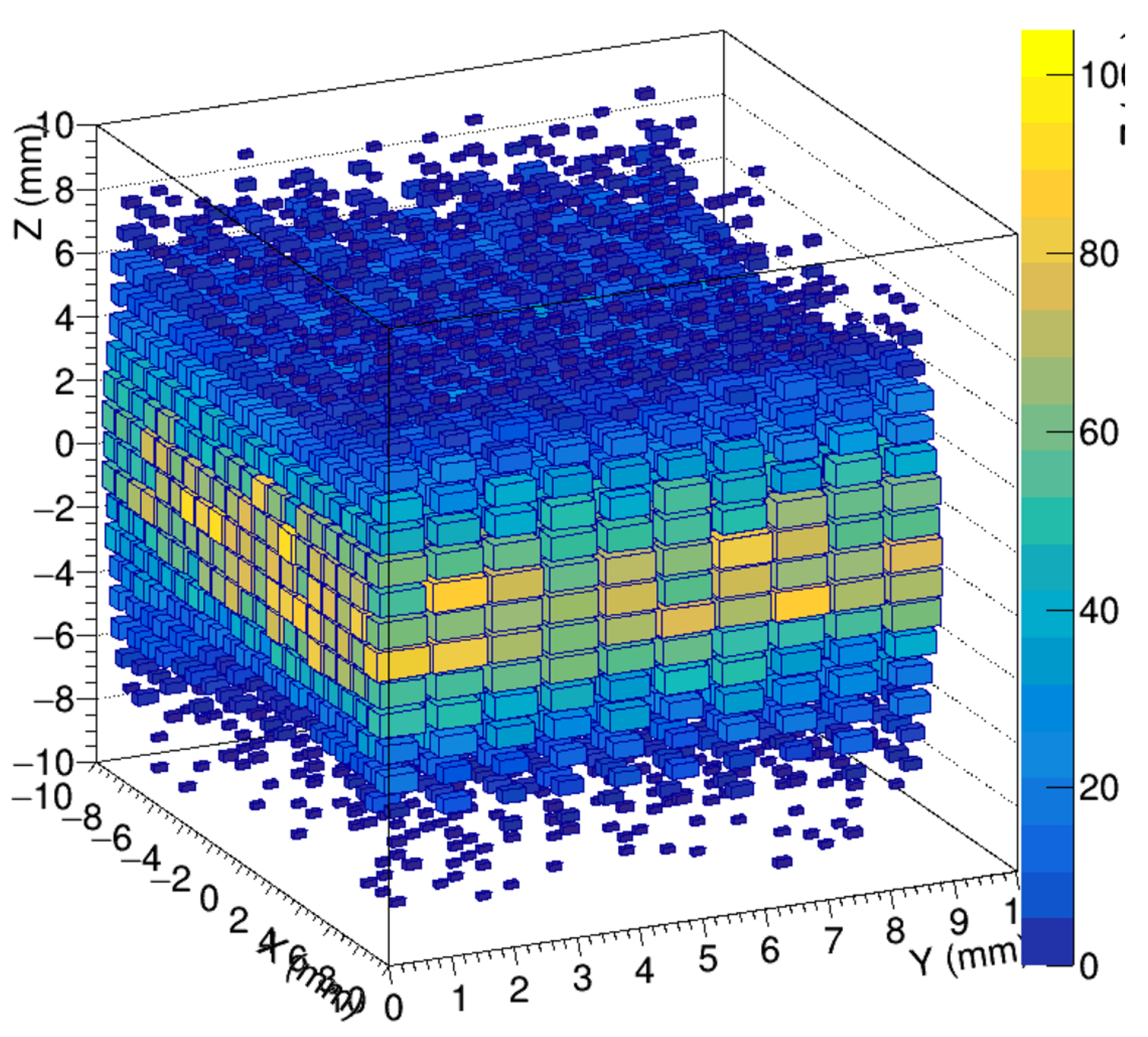}}
\end{minipage}
\begin{minipage}[h]{0.49\linewidth}
{\centering\includegraphics[scale=0.4]{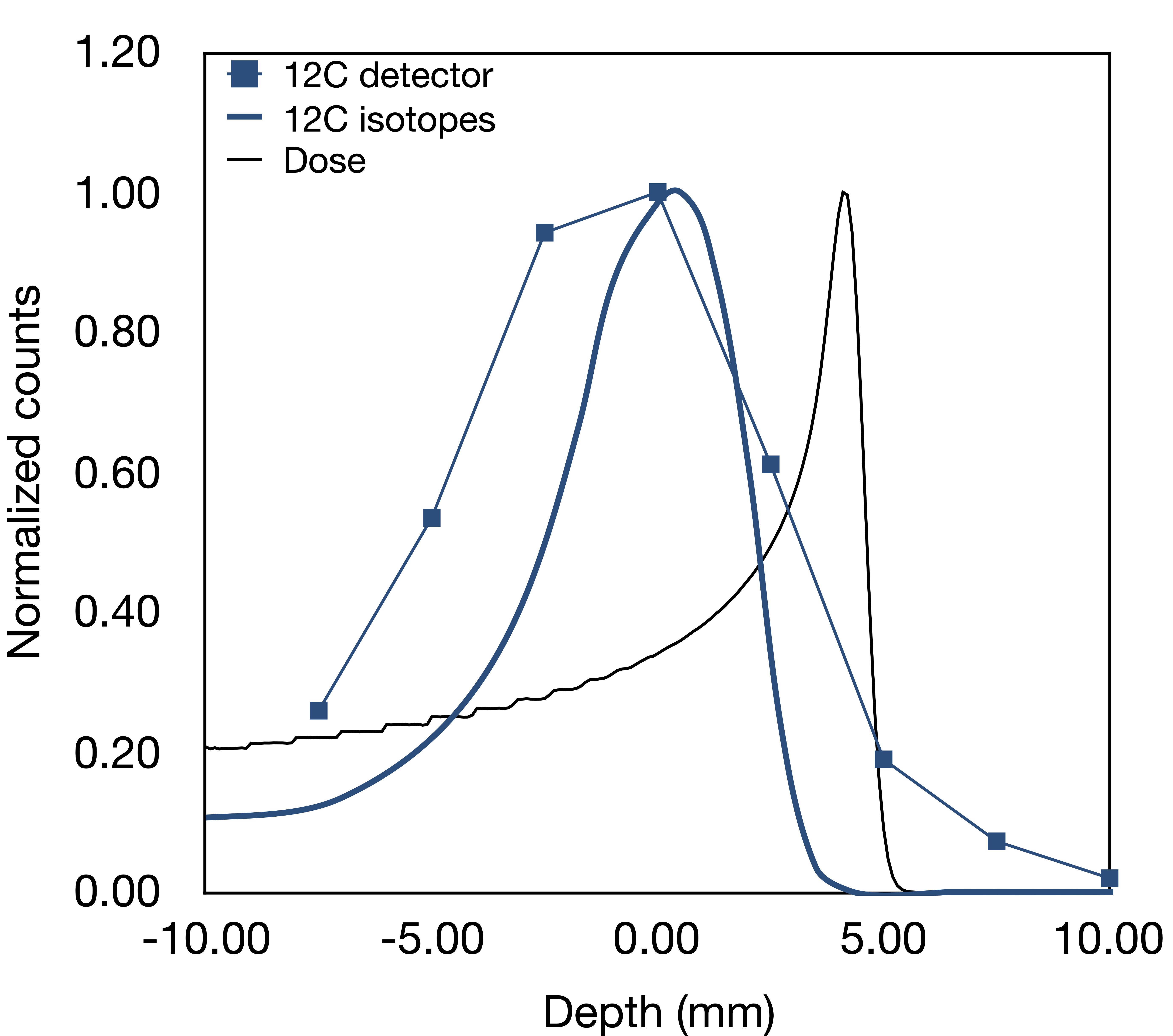}}
\end{minipage}
\caption{ Left: 3D scan of a 990 keV gamma point source to observe the simulated detector response in each axis. Right: In a simulation of a water target irradiated with 40 MeV protons, the intensity of gamma lines originating from $\mathrm{^{12}C^*}$ nuclei de-excitation as registered by eight different sectors along the detector model. Consecutive sectors are separated axially by 2.5 mm.  The original depth distribution of $\mathrm{^{12}C^*}$  is shown for comparison.}
\label{fig_3D_depth}
\end{figure}

A cubical water phantom (8 cm $\times$ 8 cm $\times$ 15 cm) was irradiated with 40 MeV protons with a range of 14.9 mm. A total of $\mathrm{1\times 10^{10}}$ protons were irradiated on the sample. Each sector observed the gamma originating from a 2.5 mm window along the beam axis. The spectrum recorded by each spectra is further analyzed to identify the PG peaks corresponding to $\mathrm{^{16}O}$ and $\mathrm{^{12}C}$ to estimate the counts. The counts from each sector contribute to the counts at a given depth along the beam axis, these are plotted in figure \ref{fig_3D_depth} (right).

Optimal strategy to position the detectors in order to  best recover the depth distribution is currently being investigated. The positions of the module and the detector design used in this simulation were in keeping with realistic clinical requirements. The separation of the front face of the detection system from the beam axis is 30 cm since the detector cannot be positioned arbitrarily close to the patient. 

Based on the simulated distribution of the detector counts it can be established that the beam currents and irradiation times utilized in a treatment setting can gather enough statistics to measure the count distributions of the gamma lines. 
When this approach is used for 3DPG application to image a smaller target sample in a  non-clinical environment, the design parameters can be adapted to achieve a much higher efficiency at a better positional resolution, along with a bigger range of positions.  A detection system with a $\mathrm{4\pi}$ coverage with 24 sectors can image 24 depth positions simultaneously.

\section{Discussion \label{Discussion}} \label{section_Discussion}

\paragraph{Knife Edge-PET}
The sum of the counts from all the crystals along a single column can be seen in figure \ref{fig_KE_depth} (a) are too few to indicate any specific features that allow the energy identification due to the small crystal size. However, when the value of the counts above the set energy threshold is plotted along different columns along the detector, there is an observable variation which can be attributed to the presence of  the knife-edge collimator which selectively favors the gamma originating from the center, and increasingly attenuates the gamma originating away from the center. As seen in figure \ref{fig_KE_depth} (b) causes a contrast of $\mathrm{4\times}$ at the peak value (column6-10) in comparison with the higher columns (20 and above) showing the unavoidable background events corresponding to the region beyond Bragg Peak.  These are due to the PG scattered by the collimator and the gamma resulting from forward neutron interactions with the phantom. Since the counts along the column are affected by the neighbouring columns due to scattering, there is a noticeable shift in the absolute peak position recorded by the detector in comparision with the expected peak position. For the nominal case of 0 mm shifted phantom with 90 MeV protons, the expected peak is at the detector center (columns 15-16) while the observed peak is 7 columns away (7$\times$3.2 mm = 22.4mm). This absolute shift is affected by the geometry, and the gamma distribution and interaction both of which can be simulated reliably to make a look up table of absolute shifts. The measured range shifts, then, will be the values corrected for these absolute shifts which is shown in figure \ref{fig_KE_depth} (c).  Phantom shifts for a given configuration between $\mathrm{\pm 10\ mm}$ can be prediceted by the detector within $\mathrm{\pm 1\ mm}$.  By implementing two such modules,  the detected PG signal can be complemented by the beamOFF 511 keV coincidence measurement by separating the collimators after beam delivery and moving the detectors closer to increase the detection efficiency. 

\paragraph{Fiducial marker  emission}
The most prominent gamma emitted by the titanium marker is the 984 keV PG which has a strong expression which stands against the background generated by water in the INER experiment as seen in figure \ref{fig_FM_INER}. The proton beam energy scan shown conducted in CGMH shows an increasing presence of the 984 keV gamma from $\mathrm{^{48}Ti}$, background corrected, increasing from 23\% to 62\% as the beam deposits the bulk of the dose at the entrance and exit of the 3 mm thin titanium marker. This shows that inserting such a marker can be a fool-proof method to verify the extent to which a proton beam traverses within the patient. In the experiment, a HPGe detector with a small crystal size and a high dead time due to the background noise contributed to limited counts. Utilizing large inorganic scintillating crystals with a fast decay time can be helpful to measure this in a treatment scenario. Alternatively, smaller crystals with photodetetors can be attached directly to the fiducial marker to achieve a higher signal. 
An important difference in our approach when compared to the test setup by \cite{Bello2020} is that the latter utilize a physical shift in the target to achieve a relative shift in the Bragg peak to the detection system, while in our case we perform an energy scan and report the results with the R80 values. 

\paragraph{3D-PG imaging}
A collimator gap of 5 mm provides nearly five times the efficiency when compared to a 1 mm collimator at the center of the collimator gap while the spatial resolution is five time worse. Enhancing the detection efficiency by combining multiple modules in a same sector enables us to add more sectors to get a larger observable region. 
The detector design presented optimizes the available space to obtain a high efficiency  and the design can easily be adapted for a non-clinical usage. In comparison to the system presented described in \cite{Verburg2018} and other spectroscopic detectors, the design presented in our work can achieve a position sensitivity for the detected gamma along the beam axis which can facilitate a 3D detection when the detector is synchronized with the beam.

\paragraph{Advantages and Disadvantages}
The PIGI imaging approaches presented in this work each have  advantages and shortcomings and are applicable in different scenarios. The knife-edge collimator in tandem with a compact PET module is an effective way to perform range verification using a compact detection system.  ASPET module together with the knife-edge collimator (4cm thick) would have a total weight of under 5 kg and can be conveniently mounted on a movable arm attached to the patient couch. Furthermore the detector can be used in a dual mode to image the PG during beamON with the collimators in position and to image the PAG during the beamOFF by separating the collimator heads using linear stages. PG detectors described for clinical applications with spectroscopic detectors by \cite{Verburg2018}, Compton camera by \cite{Draeger2018} and  knife-edge collimator based detector by \cite{Smeets2016} are best applicable when the particle beam irradiates the patient head from the top. The usage of bulky crystals and collimators that need precise alignment and dedicated stations restricts their application in regions other than the human head. Although the proposed approach is more suited for treatment sites other than the human head, there are challenges involved in precise positioning.

An insertable Ti marker is a compact and a clinically safe technique to obtain a fool-proof signal to verify the position of the maximum dose by relying on the gamma signal that is dependent on proton energy. In this implementation, the challenges on the detector are less restrictive since the position information is not of primary importance. A single crystal with a large volume and high energy resolution coupled to a PMT is sufficient to identify the characteristic PG line. Shielding from the background is less critical if the characteristic gamma signal from the chosen marker has no overlap with the typical gamma lines observed from a human body which is predominantly water. As we have found in our work, this task can be accomplished with a Ti marker although other elements are also being investigated. Insertable markers are relevant for prostrate cancer where they can be placed along with an insertable water balloon used to position the prostrate. Insertable markers are also applicable for positions close to the  tumors in head and neck, lung or the organs close to the stomach. In this case, the proton energy can be tuned to reach the marker first, and use this to verify the treatment plan indirectly. Although the net dose deposited in the sum total of all runs is unrealistic for range verification on a real patient in a single run, utilizing larger detector modules positioned closer to the patient can significantly improve the statistics.

The 3DPG detector is an optimal way to maximize the detection efficiency while retaining the ability to obtain precise positional information of the source over a small but variable region of interest. 3G-PG imaging system can be implemented not only to perform range verification in a treatment scenario but can also be used to to identify material composition by using a proton beam as a scanning probe. The X and Y coordinates of the  beam can be obtained from the delivery system and used to position the detector accordingly. The Z coordinate of the source is obtained by positioning the sectors in different regions of interest along  the beam axis. The resolution achievable by the 3DPG detector is limited by the non-uniform response of the detector to events originating radially away from the beam axis. The setup will be more suitable for non-clinical applications where space is not a constraint.

\paragraph{Challenges}
The main challenge in the described approaches for PIGI is the presence of multiple sources of background radiation. Neutron radiation is the least clearly understood sources of background as the neutron production and interaction cross sections are available in limited angles and energies. Studies to compare Geant4 and FLUKA production of secondary particles by \cite{Robert2013}  found differences between the codes in the predicted prompt gamma yields by up to 100\% , PAG up to 20\% and differences in the neutron yield and energy distributions when compared to the experiment. One of their conclusions was that the neutron production is highest in the forward region of the beam, and that the higher energy neutron flux is more likely  to be found in the forward regions. One problem with neutron background is the interaction with other elements in the test environment leading to the production of a gamma background. A potential cause of concern is the possibility of detector damage due to neutron irradiation when utilizing scintillating crystals and PET type detectors employing SiPMs. \cite{Ulyanov2020} found that a 2x2 array  of $\mathrm{6\times6\ mm^2}$ SiPMs by SensL (J-series) with an area of $\mathrm{1.47\ cm^{2}}$ experienced an increase in dark current from 100 uA for unirradiated samples to $\mathrm{5 \times 10^{4} \mu A }$ after being irradiated with a 1MeV neutron equivalent fluence of protons $\mathrm{1.28\times 10^{8} n_{eq}.cm^{-2}}$ to $\mathrm{ 1.23\times 10^{10} n_{eq}.cm^{-2}}$.  \cite{Garutti2019} reported an increase in the dark count by a factor $\mathrm{5\times 10^{3}}$ after being  irradiated by 1MeV neutron equivalent fluence of protons  $\mathrm{1 \times 10^{9}n_{eq}.cm^{-2} }$; this was subsequently down to a factor $\mathrm{1.52 \times 10^{3} }$ after annealing at 250 C for three days. In regular operation during proton therapy, the distance from the patient and the positioning of the detector can reduce significantly the effect of neutrons. The ratio of the secondary neutrons to the primary protons is energy dependent and is described in \cite{JiaNeutron2014} as 2\%  at 70 MeV and 10\% at 140 MeV based on MCNPX simulation study. Using these values we can expect the neutron flux.$\mathrm{cm^{-2}}$ to be between $\mathrm{2 \times 10^{3}\ -\ 1 \times 10^{4} \ n_{eq}.cm^{-2} }$ which is 5-6 orders of magnitude lower than the irradiation levels studied by  \cite{Garutti2019} and  \cite{Ulyanov2020}. Although crystals with a larger volume receive a higher exposure, studies by \cite{ZhangNeutron2009} with 4 MeV neutrons with a $\mathrm{6 \times 10^{8}n_{eq}.cm^{-2} }$ flux show that LYSO, BGO, CeF3  and PWO crystals do not show any noticeable  degradation in optical properties. \cite{HuNeutron2020} studied a higher  with a higher flux of $\mathrm{9\times 10^{15}\ n_{eq}.cm^{-2}}$ of 1MeV protons and conclude that LYSO/LFS crystals continue to remain radiation hard. In order to reduce the effects of neutron background  nevertheless, thick layers of PE followed by lead can be helpful. Simulations indicate that most of the produced neutrons are in the forward region, which also indicates that the detectors should be placed at 90$^o$ or higher angles from the beam axis for better safety.

Another important challenge in PIGI imaging involves the precise information of the production cross sections for gamma emitting isotopes. The models used to evaluate the production cross sections in Geant4 have previously been compared with other Monte-Carlo tool kits for high energy physics such as FLUKA and PHITS and inconsistencies were found across the predictions from different models with the experimental data \cite{ProtonTotalCrosssections}. In \cite{Wronska2021} the authors measured the  contributions of $\mathrm{^{12}C}$ and $\mathrm{^{16}O}$ PG lines and found qualitative differences in the relative values estimated by the simulation in comparison with the experiment.  We plan to measure the cross sections of the major gamma lines of interest especially in the low energy regions to address this issue.


\section{Conclusions \label{Summary and Conclusions}} \label{section_Conclusions}

Proton induced gamma imaging has potential medical applications for  range verification to enhance the scope and quality of treatment planning, identifying tissue composition and for applications in non-destructive imaging of volume samples of any material. In this work we present three approaches towards proton induced gamma imaging.
Utilizing compact detectors can make their usage in a treatment setting more practical. The proposed approaches in sections \ref{section_KE} and \ref{section_FM} can be  applied to verify the treatment of cancers related to head and neck and prostrate respectively although they can be adopted to other cases. Of special interest will be pediatric treatment where the target volumes are smaller, and the range verification is crucial for the long term survival. KE-PET approach demonstrates through simulations that PET type multi-channel detectors with smaller crystal sizes used along with knife-edge collimation with a total weight around 5-10 kg can be utilized to obtain the depth distributions of the PG. We are currently developing a PET module which can be used to evaluate this experimentally. Another approach towards range verification is to use fiducial markers made of materials such as Ti which emit a characteristic proton-induced PG distinguishable from the surrounding tissue. Our experiments have identified the characteristic gamma lines of interest and demonstrate the feasibility in a clinical environment. R80 value of the delivered beam can be verified physically by placing a marker in the region of interest within the tumor or in close proximity to it. Utilizing larger crystals placed closer to the target can provide a better information for this approach. The 3DPG detection scheme described in section \ref{section_3DPG}  has applications for non-destructive imaging of volume material samples by combining the proposed approach in this work with the information about the beam delivered to the target. By adapting the design  parameters to the volume of interest and the range proton energies desired, a 3D image with a mm level resolution can be  possible in principle. We have demonstrated a clinical application for range verification by simulating the detection system to locate the peaks of $\mathrm{^{12}C}$ PG distribution and matching it with the peak of the intrinsic isotopic distribution of the corresponding gamma.  In order to get a realistic image of the material composition it is necessary to have a reliable database of gamma production cross-sections and more studies are required on the reconstruction from detector to the gamma activity and eventually to the material distribution.

\section*{Acknowledgement}
Y. H. Wang assisted with the design and the fabrication of the mechanics and helped with the execution of the beam tests. C-C Lee, T Chao, and the medical physicists from CGMH helped organize the beam tests at Chang Gung Memorial Hospital. Dr Du helped organize the beam tests at INER along with support staff. ASGC (Academia Sinica Grid-computing Center) Distributed Cloud resources were utilized for running Monte Carlo simulations. This research was supported by the AS Thematic program grant number AS-TP-108-ML0.



\end{document}